\documentclass[longauth]{aaEC}

\usepackage{lastpage}
\usepackage{graphicx}
\usepackage{natbib}
\usepackage{scalerel}
\usepackage{csquotes}
\usepackage{tabularx}
\usepackage{booktabs}  % for better looking tables
\usepackage{siunitx}  % for better number alignment

\usepackage[table]{xcolor}

\usepackage{txfonts}
\usepackage[pdfencoding=auto,psdextra]{hyperref}
\hypersetup{
    colorlinks=true,
    linkcolor=blue,
    filecolor=magenta,      
    urlcolor=blue,
    citecolor=blue
}
\makeatletter
\renewcommand*\aa@pageof{, page \thepage{} of \pageref*{LastPage}}
\makeatother

\usepackage[utf8]{inputenc}

\usepackage[switch, modulo]{lineno}
\usepackage{amsmath,amssymb}
\usepackage{siunitx}
\usepackage{xspace}
\usepackage[nameinlink,capitalise]{cleveref}
\usepackage{euclid}
\usepackage{glossaries}
\glsdisablehyper

\newcommand{\dav}[1]{#1}

\newcommand{\davtwo}[1]{#1}

\newcommand{\davthree}[1]{#1}

\newcommand{\logT}{\ensuremath{\logten(T_{\rm AGN}/{\rm K})}}
\newcommand{\orcid}[1]{} %% if already defined in aa.cls: comment, or use renewcommand		

\defcitealias{ISTF2020}{EC20}
\defcitealias{Lacasa_2019}{LG19}
\defcitealias{Barreira2018response_approach}{B18}

\begin{document}

\title{\textit{Euclid} preparation}
\subtitle{LII. Forecast impact of super-sample covariance on 3$\times$2pt analysis with Euclid}

\author{Euclid Collaboration: D.~Sciotti\orcid{0009-0008-4519-2620}\thanks{\email{davide.sciotti@uniroma1.it}}$^{1,2,3}$,
S.~Gouyou~Beauchamps$^{4,5,6}$,
V.~F.~Cardone$^{2,3}$,
S.~Camera\orcid{0000-0003-3399-3574}$^{7,8,9}$,
I.~Tutusaus\orcid{0000-0002-3199-0399}$^{10,11,12,5}$,
F.~Lacasa\orcid{0000-0002-7268-3440}$^{11,13}$,
A.~Barreira$^{14,15}$,
M.~Bonici\orcid{0000-0002-8430-126X}$^{20}$,
A.~Gorce\orcid{0000-0002-1712-737X}$^{16}$,
M.~Aubert$^{17,18}$
% ...
}

%%%% please do not edit the affiliation list -- contact ECEB Bureau for changes
\institute{$^{1}$ Dipartimento di Fisica, Sapienza Universit\`a di Roma, Piazzale Aldo Moro 2, 00185 Roma, Italy\\
$^{2}$ INAF-Osservatorio Astronomico di Roma, Via Frascati 33, 00078 Monteporzio Catone, Italy\\
$^{3}$ INFN-Sezione di Roma, Piazzale Aldo Moro, 2 - c/o Dipartimento di Fisica, Edificio G. Marconi, 00185 Roma, Italy\\
$^{4}$ Aix-Marseille Universit\'e, CNRS/IN2P3, CPPM, Marseille, France\\
$^{5}$ Institut d'Estudis Espacials de Catalunya (IEEC),  Edifici RDIT, Campus UPC, 08860 Castelldefels, Barcelona, Spain\\
$^{6}$ Institut de Ciencies de l'Espai (IEEC-CSIC), Campus UAB, Carrer de Can Magrans, s/n Cerdanyola del Vall\'es, 08193 Barcelona, Spain\\
$^{7}$ Dipartimento di Fisica, Universit\`a degli Studi di Torino, Via P. Giuria 1, 10125 Torino, Italy\\
$^{8}$ INFN-Sezione di Torino, Via P. Giuria 1, 10125 Torino, Italy\\
$^{9}$ INAF-Osservatorio Astrofisico di Torino, Via Osservatorio 20, 10025 Pino Torinese (TO), Italy\\
$^{10}$ Institut de Recherche en Astrophysique et Plan\'etologie (IRAP), Universit\'e de Toulouse, CNRS, UPS, CNES, 14 Av. Edouard Belin, 31400 Toulouse, France\\
$^{11}$ Universit\'e de Gen\`eve, D\'epartement de Physique Th\'eorique and Centre for Astroparticle Physics, 24 quai Ernest-Ansermet, CH-1211 Gen\`eve 4, Switzerland\\
$^{12}$ Institute of Space Sciences (ICE, CSIC), Campus UAB, Carrer de Can Magrans, s/n, 08193 Barcelona, Spain\\
$^{13}$ Universit\'e Paris-Saclay, CNRS, Institut d'astrophysique spatiale, 91405, Orsay, France\\
$^{14}$ Excellence Cluster ORIGINS, Boltzmannstrasse 2, 85748 Garching, Germany\\
$^{15}$ Ludwig-Maximilians-University, Schellingstrasse 4, 80799 Munich, Germany\\
$^{16}$ Department of Physics and Trottier Space Institute, McGill University, 3600 University Street, Montreal, QC H3A 2T8, Canada\\
$^{17}$ Universit\'e Claude Bernard Lyon 1, CNRS/IN2P3, IP2I Lyon, UMR 5822, Villeurbanne, F-69100, France\\
$^{18}$ Universit\'e Clermont Auvergne, CNRS/IN2P3, LPC, F-63000 Clermont-Ferrand, France\\
$^{19}$ Jodrell Bank Centre for Astrophysics, Department of Physics and Astronomy, University of Manchester, Oxford Road, Manchester M13 9PL, UK\\
$^{20}$ INAF-IASF Milano, Via Alfonso Corti 12, 20133 Milano, Italy\\
$^{21}$ Institute for Theoretical Particle Physics and Cosmology (TTK), RWTH Aachen University, 52056 Aachen, Germany\\
$^{22}$ Universit\'e Paris-Saclay, CNRS/IN2P3, IJCLab, 91405 Orsay, France\\
$^{23}$ Centre National d'Etudes Spatiales -- Centre spatial de Toulouse, 18 avenue Edouard Belin, 31401 Toulouse Cedex 9, France\\
$^{24}$ Institut f\"ur Theoretische Physik, University of Heidelberg, Philosophenweg 16, 69120 Heidelberg, Germany\\
$^{25}$ Universit\'e St Joseph; Faculty of Sciences, Beirut, Lebanon\\
$^{26}$ Department of Astrophysics, University of Zurich, Winterthurerstrasse 190, 8057 Zurich, Switzerland\\
$^{27}$ Institute of Cosmology and Gravitation, University of Portsmouth, Portsmouth PO1 3FX, UK\\
$^{28}$ INAF-Osservatorio Astronomico di Brera, Via Brera 28, 20122 Milano, Italy\\
$^{29}$ INAF-Osservatorio di Astrofisica e Scienza dello Spazio di Bologna, Via Piero Gobetti 93/3, 40129 Bologna, Italy\\
$^{30}$ SISSA, International School for Advanced Studies, Via Bonomea 265, 34136 Trieste TS, Italy\\
$^{31}$ IFPU, Institute for Fundamental Physics of the Universe, via Beirut 2, 34151 Trieste, Italy\\
$^{32}$ INAF-Osservatorio Astronomico di Trieste, Via G. B. Tiepolo 11, 34143 Trieste, Italy\\
$^{33}$ INFN, Sezione di Trieste, Via Valerio 2, 34127 Trieste TS, Italy\\
$^{34}$ Dipartimento di Fisica e Astronomia, Universit\`a di Bologna, Via Gobetti 93/2, 40129 Bologna, Italy\\
$^{35}$ INFN-Sezione di Bologna, Viale Berti Pichat 6/2, 40127 Bologna, Italy\\
$^{36}$ Institut de Physique Th\'eorique, CEA, CNRS, Universit\'e Paris-Saclay 91191 Gif-sur-Yvette Cedex, France\\
$^{37}$ Institut d'Astrophysique de Paris, UMR 7095, CNRS, and Sorbonne Universit\'e, 98 bis boulevard Arago, 75014 Paris, France\\
$^{38}$ Dipartimento di Fisica, Universit\`a di Genova, Via Dodecaneso 33, 16146, Genova, Italy\\
$^{39}$ INFN-Sezione di Genova, Via Dodecaneso 33, 16146, Genova, Italy\\
$^{40}$ Department of Physics "E. Pancini", University Federico II, Via Cinthia 6, 80126, Napoli, Italy\\
$^{41}$ INAF-Osservatorio Astronomico di Capodimonte, Via Moiariello 16, 80131 Napoli, Italy\\
$^{42}$ Instituto de Astrof\'isica e Ci\^encias do Espa\c{c}o, Universidade do Porto, CAUP, Rua das Estrelas, PT4150-762 Porto, Portugal\\
$^{43}$ Institut de F\'{i}sica d'Altes Energies (IFAE), The Barcelona Institute of Science and Technology, Campus UAB, 08193 Bellaterra (Barcelona), Spain\\
$^{44}$ Port d'Informaci\'{o} Cient\'{i}fica, Campus UAB, C. Albareda s/n, 08193 Bellaterra (Barcelona), Spain\\
$^{45}$ Dipartimento di Fisica e Astronomia "Augusto Righi" - Alma Mater Studiorum Universit\`a di Bologna, via Piero Gobetti 93/2, 40129 Bologna, Italy\\
$^{46}$ INFN section of Naples, Via Cinthia 6, 80126, Napoli, Italy\\
$^{47}$ Dipartimento di Fisica e Astronomia "Augusto Righi" - Alma Mater Studiorum Universit\`a di Bologna, Viale Berti Pichat 6/2, 40127 Bologna, Italy\\
$^{48}$ Institut national de physique nucl\'eaire et de physique des particules, 3 rue Michel-Ange, 75794 Paris C\'edex 16, France\\
$^{49}$ Instituto de Astrof\'isica de Canarias, Calle V\'ia L\'actea s/n, 38204, San Crist\'obal de La Laguna, Tenerife, Spain\\
$^{50}$ Institute for Astronomy, University of Edinburgh, Royal Observatory, Blackford Hill, Edinburgh EH9 3HJ, UK\\
$^{51}$ European Space Agency/ESRIN, Largo Galileo Galilei 1, 00044 Frascati, Roma, Italy\\
$^{52}$ ESAC/ESA, Camino Bajo del Castillo, s/n., Urb. Villafranca del Castillo, 28692 Villanueva de la Ca\~nada, Madrid, Spain\\
$^{53}$ Institute of Physics, Laboratory of Astrophysics, Ecole Polytechnique F\'ed\'erale de Lausanne (EPFL), Observatoire de Sauverny, 1290 Versoix, Switzerland\\
$^{54}$ UCB Lyon 1, CNRS/IN2P3, IUF, IP2I Lyon, 4 rue Enrico Fermi, 69622 Villeurbanne, France\\
$^{55}$ Mullard Space Science Laboratory, University College London, Holmbury St Mary, Dorking, Surrey RH5 6NT, UK\\
$^{56}$ Departamento de F\'isica, Faculdade de Ci\^encias, Universidade de Lisboa, Edif\'icio C8, Campo Grande, PT1749-016 Lisboa, Portugal\\
$^{57}$ Instituto de Astrof\'isica e Ci\^encias do Espa\c{c}o, Faculdade de Ci\^encias, Universidade de Lisboa, Campo Grande, 1749-016 Lisboa, Portugal\\
$^{58}$ Department of Astronomy, University of Geneva, ch. d'Ecogia 16, 1290 Versoix, Switzerland\\
$^{59}$ INFN-Padova, Via Marzolo 8, 35131 Padova, Italy\\
$^{60}$ INAF-Istituto di Astrofisica e Planetologia Spaziali, via del Fosso del Cavaliere, 100, 00100 Roma, Italy\\
$^{61}$ Universit\'e Paris-Saclay, Universit\'e Paris Cit\'e, CEA, CNRS, AIM, 91191, Gif-sur-Yvette, France\\
$^{62}$ INAF-Osservatorio Astronomico di Padova, Via dell'Osservatorio 5, 35122 Padova, Italy\\
$^{63}$ Max Planck Institute for Extraterrestrial Physics, Giessenbachstr. 1, 85748 Garching, Germany\\
$^{64}$ Universit\"ats-Sternwarte M\"unchen, Fakult\"at f\"ur Physik, Ludwig-Maximilians-Universit\"at M\"unchen, Scheinerstrasse 1, 81679 M\"unchen, Germany\\
$^{65}$ Dipartimento di Fisica "Aldo Pontremoli", Universit\`a degli Studi di Milano, Via Celoria 16, 20133 Milano, Italy\\
$^{66}$ INFN-Sezione di Milano, Via Celoria 16, 20133 Milano, Italy\\
$^{67}$ Institute of Theoretical Astrophysics, University of Oslo, P.O. Box 1029 Blindern, 0315 Oslo, Norway\\
$^{68}$ Jet Propulsion Laboratory, California Institute of Technology, 4800 Oak Grove Drive, Pasadena, CA, 91109, USA\\
$^{69}$ Department of Physics, Lancaster University, Lancaster, LA1 4YB, UK\\
$^{70}$ von Hoerner \& Sulger GmbH, Schlossplatz 8, 68723 Schwetzingen, Germany\\
$^{71}$ Technical University of Denmark, Elektrovej 327, 2800 Kgs. Lyngby, Denmark\\
$^{72}$ Cosmic Dawn Center (DAWN), Denmark\\
$^{73}$ Max-Planck-Institut f\"ur Astronomie, K\"onigstuhl 17, 69117 Heidelberg, Germany\\
$^{74}$ Department of Physics and Astronomy, University College London, Gower Street, London WC1E 6BT, UK\\
$^{75}$ Department of Physics and Helsinki Institute of Physics, Gustaf H\"allstr\"omin katu 2, 00014 University of Helsinki, Finland\\
$^{76}$ Department of Physics, P.O. Box 64, 00014 University of Helsinki, Finland\\
$^{77}$ Helsinki Institute of Physics, Gustaf H{\"a}llstr{\"o}min katu 2, University of Helsinki, Helsinki, Finland\\
$^{78}$ NOVA optical infrared instrumentation group at ASTRON, Oude Hoogeveensedijk 4, 7991PD, Dwingeloo, The Netherlands\\
$^{79}$ Centre de Calcul de l'IN2P3/CNRS, 21 avenue Pierre de Coubertin 69627 Villeurbanne Cedex, France\\
$^{80}$ Universit\"at Bonn, Argelander-Institut f\"ur Astronomie, Auf dem H\"ugel 71, 53121 Bonn, Germany\\
$^{81}$ Aix-Marseille Universit\'e, CNRS, CNES, LAM, Marseille, France\\
$^{82}$ Department of Physics, Institute for Computational Cosmology, Durham University, South Road, DH1 3LE, UK\\
$^{83}$ Universit\'e C\^{o}te d'Azur, Observatoire de la C\^{o}te d'Azur, CNRS, Laboratoire Lagrange, Bd de l'Observatoire, CS 34229, 06304 Nice cedex 4, France\\
$^{84}$ Universit\'e Paris Cit\'e, CNRS, Astroparticule et Cosmologie, 75013 Paris, France\\
$^{85}$ Institut d'Astrophysique de Paris, 98bis Boulevard Arago, 75014, Paris, France\\
$^{86}$ CEA Saclay, DFR/IRFU, Service d'Astrophysique, Bat. 709, 91191 Gif-sur-Yvette, France\\
$^{87}$ European Space Agency/ESTEC, Keplerlaan 1, 2201 AZ Noordwijk, The Netherlands\\
$^{88}$ Department of Physics and Astronomy, University of Aarhus, Ny Munkegade 120, DK-8000 Aarhus C, Denmark\\
$^{89}$ Space Science Data Center, Italian Space Agency, via del Politecnico snc, 00133 Roma, Italy\\
$^{90}$ Institute of Space Science, Str. Atomistilor, nr. 409 M\u{a}gurele, Ilfov, 077125, Romania\\
$^{91}$ Departamento de Astrof\'isica, Universidad de La Laguna, 38206, La Laguna, Tenerife, Spain\\
$^{92}$ Dipartimento di Fisica e Astronomia "G. Galilei", Universit\`a di Padova, Via Marzolo 8, 35131 Padova, Italy\\
$^{93}$ Departamento de F\'isica, FCFM, Universidad de Chile, Blanco Encalada 2008, Santiago, Chile\\
$^{94}$ Centro de Investigaciones Energ\'eticas, Medioambientales y Tecnol\'ogicas (CIEMAT), Avenida Complutense 40, 28040 Madrid, Spain\\
$^{95}$ Instituto de Astrof\'isica e Ci\^encias do Espa\c{c}o, Faculdade de Ci\^encias, Universidade de Lisboa, Tapada da Ajuda, 1349-018 Lisboa, Portugal\\
$^{96}$ Universidad Polit\'ecnica de Cartagena, Departamento de Electr\'onica y Tecnolog\'ia de Computadoras,  Plaza del Hospital 1, 30202 Cartagena, Spain\\
$^{97}$ Kapteyn Astronomical Institute, University of Groningen, PO Box 800, 9700 AV Groningen, The Netherlands\\
$^{98}$ INFN-Bologna, Via Irnerio 46, 40126 Bologna, Italy\\
$^{99}$ Infrared Processing and Analysis Center, California Institute of Technology, Pasadena, CA 91125, USA\\
$^{100}$ Junia, EPA department, 41 Bd Vauban, 59800 Lille, France\\
$^{101}$ Instituto de F\'isica Te\'orica UAM-CSIC, Campus de Cantoblanco, 28049 Madrid, Spain\\
$^{102}$ CERCA/ISO, Department of Physics, Case Western Reserve University, 10900 Euclid Avenue, Cleveland, OH 44106, USA\\
$^{103}$ Laboratoire de Physique de l'\'Ecole Normale Sup\'erieure, ENS, Universit\'e PSL, CNRS, Sorbonne Universit\'e, 75005 Paris, France\\
$^{104}$ Observatoire de Paris, Universit\'e PSL, Sorbonne Universit\'e, LERMA, 750 Paris, France\\
$^{105}$ Astrophysics Group, Blackett Laboratory, Imperial College London, London SW7 2AZ, UK\\
$^{106}$ Scuola Normale Superiore, Piazza dei Cavalieri 7, 56126 Pisa, Italy\\
$^{107}$ Dipartimento di Fisica e Scienze della Terra, Universit\`a degli Studi di Ferrara, Via Giuseppe Saragat 1, 44122 Ferrara, Italy\\
$^{108}$ Istituto Nazionale di Fisica Nucleare, Sezione di Ferrara, Via Giuseppe Saragat 1, 44122 Ferrara, Italy\\
$^{109}$ Dipartimento di Fisica - Sezione di Astronomia, Universit\`a di Trieste, Via Tiepolo 11, 34131 Trieste, Italy\\
$^{110}$ NASA Ames Research Center, Moffett Field, CA 94035, USA\\
$^{111}$ Kavli Institute for Particle Astrophysics \& Cosmology (KIPAC), Stanford University, Stanford, CA 94305, USA\\
$^{112}$ INAF, Istituto di Radioastronomia, Via Piero Gobetti 101, 40129 Bologna, Italy\\
$^{113}$ Institute Lorentz, Leiden University, Niels Bohrweg 2, 2333 CA Leiden, The Netherlands\\
$^{114}$ Institute for Astronomy, University of Hawaii, 2680 Woodlawn Drive, Honolulu, HI 96822, USA\\
$^{115}$ Department of Physics \& Astronomy, University of California Irvine, Irvine CA 92697, USA\\
$^{116}$ Department of Astronomy \& Physics and Institute for Computational Astrophysics, Saint Mary's University, 923 Robie Street, Halifax, Nova Scotia, B3H 3C3, Canada\\
$^{117}$ Departamento F\'isica Aplicada, Universidad Polit\'ecnica de Cartagena, Campus Muralla del Mar, 30202 Cartagena, Murcia, Spain\\
$^{118}$ Dipartimento di Fisica, Universit\`a degli studi di Genova, and INFN-Sezione di Genova, via Dodecaneso 33, 16146, Genova, Italy\\
$^{119}$ Department of Computer Science, Aalto University, PO Box 15400, Espoo, FI-00 076, Finland\\
$^{120}$ Ruhr University Bochum, Faculty of Physics and Astronomy, Astronomical Institute (AIRUB), German Centre for Cosmological Lensing (GCCL), 44780 Bochum, Germany\\
$^{121}$ Caltech/IPAC, 1200 E. California Blvd., Pasadena, CA 91125, USA\\
$^{122}$ Department of Physics and Astronomy, Vesilinnantie 5, 20014 University of Turku, Finland\\
$^{123}$ Serco for European Space Agency (ESA), Camino bajo del Castillo, s/n, Urbanizacion Villafranca del Castillo, Villanueva de la Ca\~nada, 28692 Madrid, Spain\\
$^{124}$ Oskar Klein Centre for Cosmoparticle Physics, Department of Physics, Stockholm University, Stockholm, SE-106 91, Sweden\\
$^{125}$ Univ. Grenoble Alpes, CNRS, Grenoble INP, LPSC-IN2P3, 53, Avenue des Martyrs, 38000, Grenoble, France\\
$^{126}$ Centro de Astrof\'{\i}sica da Universidade do Porto, Rua das Estrelas, 4150-762 Porto, Portugal\\
$^{127}$ Dipartimento di Fisica, Universit\`a di Roma Tor Vergata, Via della Ricerca Scientifica 1, Roma, Italy\\
$^{128}$ INFN, Sezione di Roma 2, Via della Ricerca Scientifica 1, Roma, Italy\\
$^{129}$ Department of Mathematics and Physics E. De Giorgi, University of Salento, Via per Arnesano, CP-I93, 73100, Lecce, Italy\\
$^{130}$ INAF-Sezione di Lecce, c/o Dipartimento Matematica e Fisica, Via per Arnesano, 73100, Lecce, Italy\\
$^{131}$ INFN, Sezione di Lecce, Via per Arnesano, CP-193, 73100, Lecce, Italy\\
$^{132}$ Higgs Centre for Theoretical Physics, School of Physics and Astronomy, The University of Edinburgh, Edinburgh EH9 3FD, UK\\
$^{133}$ Department of Astrophysical Sciences, Peyton Hall, Princeton University, Princeton, NJ 08544, USA\\
$^{134}$ Niels Bohr Institute, University of Copenhagen, Jagtvej 128, 2200 Copenhagen, Denmark}  

\date{\today}

%%%%%%%%%%%%%%%%%%%%%%%%%%%%%%%%%%%%%%%%%%%%%%%%%%%%%%%%%%%%%%%%%%%%%%%

% \abstract{}{}{}{}{} 
% 5 {} token are mandatory

\abstract
{Deviations from Gaussianity in the distribution of the fields probed by large-scale structure surveys generate additional terms in the data covariance matrix, increasing the uncertainties in the measurement of the cosmological parameters. Super-sample covariance (SSC) is among the largest of these non-Gaussian contributions, with the potential to significantly degrade constraints on some of the parameters of the cosmological model under study -- especially for weak-lensing cosmic shear.}
{We compute and validate the impact of SSC on the forecast uncertainties on the cosmological parameters for the \Euclid photometric survey, and investigate how its impact depends on the specific details of the forecast.}
{We followed the recipes outlined by the Euclid Collaboration (EC) to produce $1\sigma$ constraints through a Fisher matrix analysis, considering the Gaussian covariance alone and adding the SSC term, which is computed through the public code {\tt PySSC}. The constraints are produced both by using \Euclid's photometric probes in isolation and by combining them in the `3$\times$2pt' analysis.} 
{We meet EC requirements on the forecasts validation, with an agreement at the 10\% level between the mean results of the two pipelines considered, and find the SSC impact to be non-negligible -- halving the figure of merit (FoM) of the dark energy parameters ($w_0$, $w_a$) in the 3$\times$2pt case and substantially increasing the uncertainties on $\Omega_{{\rm m},0}, w_0$, and $\sigma_8$ for the weak-lensing probe. 
We find photometric galaxy clustering to be less affected as a consequence of the lower probe response. 
\dav{The relative impact of SSC, while highly dependent on the number and type of nuisance parameters varied in the analysis, does not show significant changes under variations of the redshift binning scheme}.
Finally, we explore how the use of prior information on the shear and galaxy bias changes the  impact of SSC. We find that improving shear bias priors has no significant influence, while galaxy bias must be calibrated to a subpercent level in order to increase the FoM by the large amount needed to achieve the value when SSC is not included.}
{}


\authorrunning{D. Sciotti et al.}
\titlerunning{\Euclid preparation. LII. Forecast impact of super-sample covariance on 3$\times$2pt analysis with \Euclid}

\maketitle

\section{Introduction} \label{sec:intro}
Over recent decades, we have witnessed a remarkable improvement in the precision of cosmological experiments, and consequently in our grasp of the general properties of the Universe. The $\Lambda$ cold dark matter (CDM) concordance cosmological model provides an exquisite fit to observational data from both the very early and the very late Universe, but despite its success, the basic components it postulates are poorly understood. Moreover, the nature of the mechanism responsible for the observed accelerated cosmic expansion \citep{Riess1998, Perlmutter1999} and that of the component accounting for the vast majority of the matter content, dark matter, are still unknown.
Upcoming Stage IV surveys like the {\it Vera C. Rubin} Observatory Legacy Survey of Space and Time \citep[LSST,][]{ivezic2018lsst}, the {\it Nancy Grace Roman} Space Telescope \citep{spergel2015widefield}, and the \Euclid mission \dav{\citep{laureijs2011euclid, Mellier2024}} promise to help deepen our understanding of these dark components and the nature of gravity on cosmological scales by providing unprecedented observations of the large-scale structures (LSS) of the Universe.

Because of their high accuracy and precision, these next-generation experiments will require accurate modelling of both the theory and the covariance of the observables under study in order to produce precise and unbiased estimates of the cosmological parameters. Amongst the different theoretical issues to deal with is super-sample covariance (SSC), a form of sample variance arising from the finiteness of the survey area. SSC was first introduced for cluster counts by \citet{Hu2003}, and is sometimes referred to as `beat coupling' \citep{Rimes2006, Hamilton2006}. In recent years, SSC has received a lot of attention \citep{Takada2013, Li2014, Barreira2018response_approach, Digman2019, Bayer2022, Yao2023}; see also \citet{Linke2023} for an insightful discussion on SSC in real space. Hereafter, \citet{Barreira2018response_approach} is cited as \citetalias{Barreira2018response_approach}.

The effect arises from the coupling between `supersurvey' modes ---with wavelength $\lambda$ larger than the survey typical size $L = V_{\rm s}^{1/3}$ (where $V_{\rm s}$ is the volume of the survey)--- and short-wavelength ($\lambda<L$) modes. This coupling is in turn due to the significant non-linear evolution undergone by low-redshift cosmological probes (contrary to, for example, the cosmic microwave background), which breaks the initial homogeneity of the density field, making its growth position dependent. In Fourier space, this means that modes with different wavenumber $k =  2\pi/\lambda$ become coupled.
The modulation induced by the supersurvey modes is equivalent to a change in the background density of the observed region, which affects and correlates all LSS probes. It is accounted for as an additional, non-diagonal term in the data covariance matrix beyond the Gaussian covariance, which is the only term that would exist if the random field under study were Gaussian. Being the most affected by non-linear dynamics, the smaller scales are heavily impacted by SSC, where the effect is expected to be the dominant source of statistical uncertainty for the two-point statistics of weak-lensing cosmic shear (WL): it has in fact been found to increase conditional uncertainties by up to a factor of about 2
\citep[for a \Euclid-like survey, see ][]{Barreira2018cosmic_shear, beauchamps2021}. In the case of photometric galaxy clustering (GCph; again, for a \Euclid-like survey), \citet{Lacasa_2019} -- hereafter \citetalias{Lacasa_2019} -- found the cumulative signal-to-noise ratio to be decreased by a factor of around $6$ at $\ell_{\rm max} = 2000$.
% it used to be " $\sim 5.7 \%$ "
These works, however, either do not take into account marginalised uncertainties or the variability of the probe responses, do not include cross-correlations between probes, or \davthree{do not follow the full specifics (such as modelling of the observables, types of systematics included, binning schemes, sky coverage and so forth)} of the \Euclid survey detailed below. 

There are two aims to the present study. First, we intend to validate the forecast constraints on the cosmological parameters, both including and neglecting the SSC term; these are produced using two independent codes, whose only shared feature is their use of the public \texttt{Python} module \texttt{PySSC}\footnote{\texttt{\url{https://github.com/fabienlacasa/PySSC}}}\footnote{\texttt{\url{https://pyssc.readthedocs.io/en/latest/index.html}}} (\citetalias{Lacasa_2019}) to compute the fundamental elements needed to build the SSC matrix. Second, we investigate the impact of SSC on the marginalised uncertainties and the dark energy figure of merit (FoM), both of which are obtained through a Fisher forecast of the constraining power of \Euclid's photometric observables.

The article is organised as follows: Sect.~\ref{sec: theory} presents an overview of the SSC and the approximations used to compute it. In  Sect.~\ref{sec: specifics} we outline the theoretical model and specifics used to produce the forecasts, while Sect.~\ref{sec: validation} provides technical details regarding the implementation and validation of the code. In Sect.~\ref{sec: impact} we then present a study of the impact of SSC on \Euclid constraints for different binning schemes and choices of systematic errors and priors. Finally, we present our conclusions in Sect.~\ref{sec: conclu}.

%%%%%%%%%%%%%%%%%%%%%%%%%%%%%%%%%%%%%%%%%%%%%%%%%%%%%%%%%%%%%%%%%%%%%%%

\section{SSC theory and approximations}\label{sec: theory}

\subsection{General formalism}
Throughout the article, we work with 2D-projected observables, namely the angular Power Spectrum (PS), which in the Limber approximation \citep{Limber1953, Kaiser1998} can be expressed as
\begin{equation}\label{eq: CijSSC}
    C^{AB}_{ij}(\ell) = \int \diff V \; W^A_i(z) W^B_j(z) P_{AB}(k_\ell , z) \; ,
\end{equation}
giving the correlation between probes $A$ and $B$ in the redshift bins $i$ and $j$, as a function of the multipole $\ell$; $k_\ell = (\ell+1/2)/r(z)$ is the Limber wavenumber and $W_i^A(z), W_j^B(z)$ are the survey weight functions (WFs), or \enquote{kernels}. Here we consider as the element of integration $\diff V = r^2(z) \frac{\diff r}{\diff z}  \diff z$ which is the comoving volume element per steradian, with $r(z)$ being the comoving distance.

The SSC between two projected observables arises because real observations of the Universe are always limited by a survey window function $\mathcal{M}(\vec{x})$. Taking $\mathcal{M}(\vec{x})$ at a given redshift, thus considering only its angular dependence $\mathcal{M}(\hat{\vec{n}})$\footnote{Here we do not consider a redshift dependence of $\mathcal{M}(\hat{\vec{n}})$ but this can happen for
surveys with significant depth variations across the sky. This is discussed in \citet{Lima2018}.}, with $\hat{\vec{n}}$ the unit vector on a sphere, we can define the background density contrast as \citep{Lima2018}
\begin{equation}
    \delta_{\rm b}(z) = \frac{1}{\Omega_{\rm S}}\int \diff^2\hat n \, \mathcal{M}(\hat{\vec{n}}) \, \, \delta_{\rm m} \left[r(z) \hat{\vec{n}}, z\right] \; ,
    \label{eq:delta_b_masked}
\end{equation}
with $r(z)\hat{\vec{n}} = \vec{x}$. In this equation,  $\delta_{\rm m}(\vec{x}, z) = \left[\rho_{\rm m}(\vec{x}, z)/\bar{\rho}_{\rm m}(z)-1\right]$ is the matter density contrast, with $\rho_{\rm m}(\vec{x}, z)$ the matter density and $\bar\rho_{\rm m}(z)$ its spatial average over the whole Universe at \dav{redshift} $z$ and $\Omega_{\rm S}$ the solid angle observed by the survey. 

In other words, $\delta_{\rm b}$ is the spatial average of the density contrast $\delta_{\rm m}(\vec{x}, z)$ over the survey area:
\begin{align}
 &\langle\delta_{\rm m}(\vec{x}, z)\rangle_{\text{universe}} = 0 \; , \\
 &\langle\delta_{\rm m}(\vec{x}, z)\rangle_{\text{survey}} = \delta_{\rm b}(z)\; .
\end{align}
The covariance of this background density contrast is defined as $\sigma^2(z_1,z_2) \equiv \langle \delta_{\rm b}(z_1) \, \delta_{\rm b}(z_2) \rangle$  and in the full-sky approximation is given by \citep{LacasaRosenfeld2016}
\begin{equation}\label{eq: sigma}
   \sigma^2(z_{1}, z_{2}) = \frac{1}{2 \pi^{2}} \int \diff k \; k^{2} \,P_{\rm mm}^{\, \rm lin}\left(k , z_{12}\right)\, {\rm j}_{0}\left(k r_{1}\right)\, {\rm j}_{0}\left(k r_{2}\right) \; ,
\end{equation}
with $P_{\rm mm}^{\, \rm lin}(k , z_{12})\equiv D(z_1)\,D(z_2)\,P_{\rm mm}^{\, \rm lin}(k, z=0)$ the \textit{linear} matter cross-spectrum between $z_1$ and $z_2$, $D(z)$ the linear growth factor and ${\rm j_0}(kr_i)$ the first-order spherical Bessel function, and $r_{i}=r(z_i)$. The use of the linear PS reflects the fact that the SSC is caused by long-wavelength perturbations, which are well described by linear theory. We note that we have absorbed the $\Omega_\mathrm{S}^{-1}$ prefactor of Eq.~\eqref{eq:delta_b_masked}, equal to $4\pi$ in full sky, in the $\diff V_i$ terms, being them the comoving volume element per steradian.

Depending on the portion of the Universe observed, $\delta_{\rm b}$ will be different, and in turn the PS of the considered observables $P_{AB}(k_\ell , z)$ (appearing in Eq.~\ref{eq: CijSSC}) will react to this change in the background density through the \textit{probe response} $\partial P_{AB}(k_{\ell}, z) / \partial \delta_{\rm b}$.

SSC is then the combination of these two elements, encapsulating the covariance of $\delta_{\rm b}$ and the response of the observables to a change in $\delta_{\rm b}$; the general expression of the SSC between two projected observables is \citep{LacasaRosenfeld2016, Schaan2014, Takada2013}:
\begin{multline}\label{eq: covSSCintermediate}
   {\rm Cov_{SSC}}\left[C^{AB}_{ij}(\ell),C^{CD}_{kl}(\ell')\right]=
   \int \diff V_1 \diff V_2 \; W^A_i(z_1)\,W^B_j(z_1)\, \\
   \times W^C_k(z_2)\, W^D_l(z_2) 
   \frac{\partial P_{AB}(k_\ell , z_1)}{\partial \delta_{\rm b}}\,
   \frac{\partial P_{CD}(k_{\ell'} , z_2)}{\partial \delta_{\rm b}}\,
   \sigma^2(z_1, z_2) \; .
\end{multline}

We adopt the approximation presented in \citet{Lacasa_2019}, which assumes the responses to vary slowly in redshift with respect to $\sigma^2(z_1, z_2)$. We can then approximate the responses with their weighted average over the $W^A_i(z)$ kernels \citep{beauchamps2021}:
\begin{equation}\label{eq: avg_response}
    \frac{\partial \bar{P}_{AB}(k_{\ell}, z)}{\partial \delta_{\rm b}}
    = \frac{ \int \diff V \; W_{i}^{A}(z) W_{j}^{B}(z)\, \partial P_{AB}(k_{\ell}, z) / \partial \delta_{\rm b}}{\int \diff V \; W^A_i(z) W^B_j(z)} \; ,
\end{equation}
and pull them out of the integral. The denominator on the right-hand side (r.h.s.) acts as a normalisation term, which we call $I^{AB}_{ij}$. We can further manipulate the above expression by factorising the probe response as 
\begin{equation}\label{eq: response}
    \frac{\partial P_{AB}(k_\ell , z)}{\partial \delta_{\rm b}} = 
    R^{AB}(k_\ell, z) P_{AB}(k_\ell , z) \; ,
\end{equation}
where $R^{AB}(k_\ell, z)$, the \enquote{response coefficient}, can be obtained from simulations, as in \citet{Wagner2014, Wagner2015, Li2015, Barreira2019}, or from theory (e.g. via the halo model) as in \citet{Takada2013, Krause2017, Rizzato2019}. Following \citetalias{Lacasa_2019}, we can introduce the probe response of the angular power spectrum $C^{AB}_{ij}(\ell)$ in a similar way, using Eq.~\eqref{eq: CijSSC}
\begin{align}\label{eq: 2dresponses}
     \frac{\partial C^{AB}_{ij}(\ell)}{\partial \delta_{\rm b}} &= \int \diff V \; W_{i}^{A}(z) W_{j}^{B}(z)\, \frac{\partial P_{AB} (k_{\ell}, z)}{ \partial \delta_{\rm b}} \nonumber \\
     &\equiv R^{AB}_{ij}(\ell) C^{AB}_{ij}(\ell) \; .
\end{align}
Substituting Eq.~\eqref{eq: response} into the r.h.s. of Eq.~\eqref{eq: avg_response}, using Eq.~\eqref{eq: 2dresponses} and dividing by the sky fraction observed by the telescope $f_{\rm sky} = \Omega_\mathrm{S}/4\pi$, we get the expression of the SSC which will be used throughout this work:
\begin{multline} \label{eq: covssc}
    {\rm Cov_{SSC}}\left[C^{AB}_{ij}(\ell)\,C^{CD}_{kl}(\ell')\right] \simeq 
    f_{\rm sky}^{-1}\left[ R^{AB}_{ij}(\ell)\,C^{AB}_{ij}(\ell)\right. \\
    \times
    \left. R^{CD}_{kl}(\ell')\, C^{CD}_{kl}(\ell')\, S^{A,B;C,D}_{i,j;k,l}\right] \; .
\end{multline}
In the above equation, we define
\begin{equation}\label{eq: sijkl}
        S^{A,B;C,D}_{i,j;k,l} \equiv \int \diff V_1 \diff V_2 \; \dfrac{W^A_i(z_1)W^B_j(z_1)}{I^{AB}_{ij}} \, 
        \dfrac{W^C_k(z_2)W^D_l(z_2)}{I^{CD}_{kl}}\, \sigma^2(z_1, z_2) \, .
\end{equation}
The $S^{A,B;C,D}_{i,j;k,l}$ matrix (referred to as $S_{ijkl}$ from here on) is the volume average of $\sigma^2(z_1, z_2)$, and is a dimensionless quantity. It is computed through the public \texttt{Python} module \texttt{PySSC}, released alongside the above-mentioned \citetalias{Lacasa_2019}. A description of the way this code has been used, and some comments on the inputs to provide and the outputs it produces, can be found in Sect.~\ref{sec: validation}. 

The validity of Eq.~\eqref{eq: covssc} has been tested in \citetalias{Lacasa_2019} in the case of GCph and found to reproduce the Fisher matrix \citep[FM, ][]{tegmark1997} elements and signal-to-noise ratio from the original expression (Eq.~\ref{eq: covSSCintermediate}):
\begin{enumerate}[-]
    \item within 10\% discrepancy up to $\ell \simeq 1000$ for $R^{AB}_{ij}(k_\ell, z) = {\rm const}$; 
    \item within 5\% discrepancy up to $\ell \simeq 2000$ when using the linear approximation in scale for $R^{AB}(k_\ell, z)$ provided in Appendix C of the same work.
\end{enumerate}
The necessity to push the analysis to smaller scales, as well as to investigate the SSC impact not only for GCph but also for WL and their cross-correlation, has motivated a more exhaustive characterisation of the probe response functions, which will be detailed in the next section.\\
Another approximation used in the literature has been presented in \citep{Krause2017}: the $\sigma^2(z_1, z_2)$ term is considered as a Dirac delta in $z_1 = z_2$. This greatly simplifies the computation, because the double redshift integral $\diff V_1 \diff V_2$ collapses to a single one. This approximation is used by the other two available public codes which can compute the SSC: \texttt{PyCCL} \citep{pyccl2019} and \texttt{CosmoLike} \citep{Krause2017}. \citet{Lima2018} compared this approximation against the one used in this work, finding the former to fare better for wide redshift bins (as in the case of WL), and the latter for narrow bins (as in the case of GCph).

Lastly, we note that in Eq.~\eqref{eq: covssc} we account for the sky coverage of the survey through the full-sky approximation by simply dividing by $f_{\rm sky}$; in the case of \Euclid we have $\Omega_\mathrm{S} = 14\,700 \ \rm{deg}^2 \simeq 4.4776 \ \rm{sr}$ \dav{, which corresponds to $f_{\rm sky} \simeq 0.356$}. The validity of this approximation has been discussed in \citet{beauchamps2021}, and found to agree at the percent level on the marginalised parameter constraints with the more rigorous treatment accounting for the exact survey geometry when considering large survey areas. For this test, they considered an area of $15\,000 \ \rm{deg}^2$ and a survey geometry very close to what \Euclid will have, i.e. the full sky with the ecliptic and galactic plane removed. Intuitively, the severity of the SSC decays as $f_{\rm sky}^{-1}$ because larger survey volumes can accommodate more Fourier modes. 

We note that we are considering here the maximum sky coverage that \Euclid will reach, i.e. the final data release (DR3). For the first data release (DR1), the sky coverage will be significantly lower and the full-sky approximation will not hold. In that case, the partial-sky recipe proposed in \citet{beauchamps2021} should be considered instead.

\subsection{Probe response}\label{sec: probe_response}
As mentioned in the previous section, one of the key ingredients of the SSC is the probe response. To compute this term for the probes of interest, we build upon previous works \citep[][\citetalias{Barreira2018response_approach}]{Wagner2014, Wagner2015, Li2015, Barreira2017}, and compute the response coefficient of the matter PS as
\begin{equation}\label{eq: R1mm}
    R^{\rm mm}(k,z) = \frac{\partial \ln P_{\rm mm}(k,z)}{\partial \delta_{\rm b}} = 1  - \frac{1}{3}\frac{\partial \ln P_{\rm mm}(k,z)}{\partial \ln k} + G_1^{\rm mm}(k, z) \; ,
\end{equation}
where $G_1^{\rm mm}(k, z)$ is called the growth-only response, and is constant and equal to 26/21 in the linear regime and can be computed in the non-linear regime using separate universe simulations, as done in \citet{Wagner2015}, whose results were used in \citetalias{Barreira2018response_approach} (and in the present work). The latter uses a power law to extrapolate the values of the response for $k > k_{\rm max}$, with $k_{\rm max}$ being the maximum wavenumber at which the power spectrum is reliably measured from the simulations. Further details on this extrapolation, as well as on the redshift and scale dependence of $R^{\rm mm}$, can be found respectively in Sect.~2 and the left panel of Fig.~1 of \citetalias{Barreira2018response_approach}. We note that $R^{\rm mm}$ is the response coefficient of isotropic large-scale \textit{density} perturbations; we neglect the contribution from the anisotropic \textit{tidal-field} perturbations to the total response of the power spectrum (and consequently to the SSC), which has been shown in \citetalias{Barreira2018response_approach} to be subdominant for WL with respect to the first contribution (about 5\% of the total covariance matrix at $\ell \gtrsim 300$). While we do not expect this conclusion to change substantially for GCph, we leave an accurate assessment for future work.
    
\begin{figure}[h!]
\includegraphics[width=\hsize]{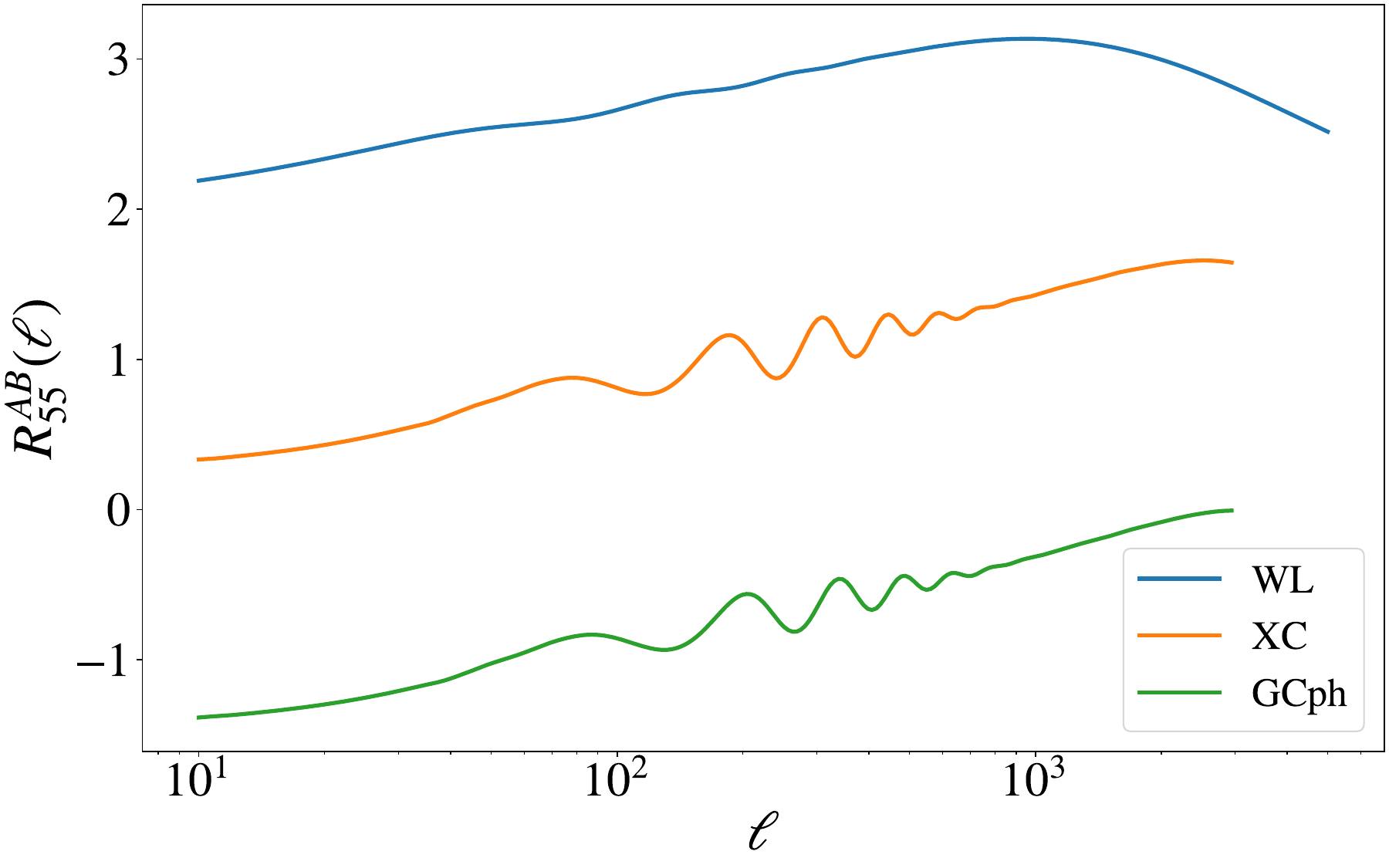}
\caption{Projected response coefficients for the WL and GCph probes and their cross-correlation for the central redshift bin ($0.8 \lesssim z \lesssim 0.9$). The shape and amplitude of the functions for different redshift pairs are analogous. For WL, the baryon acoustic oscillation wiggles are smoothed out by the projection, because the kernels are larger than the GCph ones. The different amplitude of the response is one of the main factors governing the severity of SSC.}
\label{fig: responsecoefficients}
\end{figure}

The probes considered in the present study are WL, GCph and their cross-correlation (XC); \dav{assuming general relativity}, the corresponding power spectra are given by the following expressions
\begin{equation}\label{eq: powspec}
P_{AB}(k, z) = \left \{
\begin{array}{ll}
\displaystyle{P_{\rm mm}(k, z)} & \displaystyle{A = B = {\rm L}} \\
 & \\
\displaystyle{b_{(1)}(z) P_{\rm mm}(k, z)} & \displaystyle{A = {\rm L} \, , \ \ B = {\rm G}} \\
 & \\
\displaystyle{b^2_{(1)}(z) P_{\rm mm}(k, z)} & \displaystyle{A = B = {\rm G},} \\
\end{array}
\right .
\end{equation}
with $({\rm L, G})$ for (shear, position), $P_{\rm mm}(k, z)$ the \textit{non-linear} matter PS and $b_{(1)}(z)$ the linear, scale-independent and deterministic galaxy bias. A comment is in order about the way we model the galaxy-matter and galaxy-galaxy power spectra. We are indeed using a linear bias, but the non-linear recipe for the matter power spectrum $P_{\rm mm}(k, z)$. This is reminiscent of the hybrid 1-loop perturbation theory (PT) model adopted by, for example, the DES Collaboration in the analysis of the latest data release \citep{Krause21,Pandey22}, but we drop the higher-order bias terms. This simplified model has been chosen in order to be consistent with the IST:F 
\citep[][from hereon \citetalias{ISTF2020}]{ISTF2020} forecasts, against which we compare our results (in the Gaussian case) to validate them. We are well aware that scale cuts should be performed to avoid biasing the constraints, but we are here more interested in the relative impact of SSC on the constraints than the constraints themselves. Any systematic error due to the approximate modelling should roughly cancel out in the ratio we compute later on.
We note also that we choose to include a perfectly Poissonian shot noise term in the covariance matrix, rather than in the signal, as can be seen in Eq.~\eqref{eq: noiseps}. The responses for the different probes can be obtained in terms\footnote{Since we are using the non-linear matter power spectrum $P_{\rm mm}(k, z)$, we do not force $R^{\rm mm}(k, z)$ to reduce to its linear expression, that is to say, we do not set $G_{1}^{\rm mm} = 26/21$ in Eq.~\eqref{eq: R1mm}.} of $R^{\rm mm}(k, z)$ by using the relations between matter and galaxy PS given above
\begin{equation}\label{eq: Rgg}
    R^{\rm gg}(k, z) = \frac{\partial \ln P_{\rm gg}(k,z)}{\partial \delta_{\rm b}} = R^{\rm mm}(k, z) + 2b_{(1)}^{-1}(z) \left[ b_{(2)}(z) - b_{(1)}^2(z) \right],
\end{equation}
and similarly for $R^{\rm gm}$:
\begin{equation}\label{eq: Rgm}
    R^{\rm gm}(k, z) = \frac{\partial \ln P_{\rm gm}(k,z)}{\partial \delta_{\rm b}} = R^{\rm mm}(k, z) + b_{(1)}^{-1}(z)\left[ b_{(2)}(z) - b_{(1)}^2(z) \right].
\end{equation}
Having used the definitions of the first- and second-order galaxy bias, that is, $b_{(1)}(z) = (\partial n_{\rm g} / \partial \delta_{\rm b})/n_{\rm g}$ and $b_{(2)}(z) = (\partial^2 n_{\rm g} / \partial \delta_{\rm b}^2)/n_{\rm g}$, with $n_{\rm g}$ the total angular galaxy number density, in ${\rm arcmin}^{-2}$. In the following, where there is no risk of ambiguity, we drop the subscript in parenthesis when referring to the first-order galaxy bias -- that is, $b(z) = b_{(1)}(z)$ -- to shorten the notation, and we indicate the value of the first-order galaxy bias in the $i$-th redshift bin with $b_i(z)$. More details on the computation of these terms can be found in Sect.~\ref{sec: bias}. 
We note that Eqs.~\eqref{eq: Rgg}--\eqref{eq: Rgm} are obtained by differentiating a PS model for a galaxy density contrast defined with respect to (w.r.t.) the \textit{observed} galaxy number density, and so they already account for the fact that the latter also ``responds'' to the large-scale perturbation $\delta_{\rm b}$. This is also the reason why $R^{\rm GG}_{ij}(\ell)$ can have negative values: for galaxy clustering, the (number) density contrast $\delta_{\rm gal}$ is measured w.r.t. the observed, local number density ${\bar n}_{\rm gal}$: $\delta_{\rm gal} = n_{\rm gal}/{\bar n}_{\rm gal} - 1$. The latter also responds to a background density perturbation $\delta_{\rm b}$, and it can indeed happen that ${\bar n}_{\rm gal}$ grows with $\delta_{\rm b}$ faster than $n_{\rm gal}$, which leads to $\delta_{\rm gal}$ decreasing with increasing $\delta_{\rm b}$ (which also implies $\partial C^{\rm GG}_{ij}(\ell) / \partial \delta_{\rm b} < 0$).
We also stress the fact that the second-order galaxy bias appearing in the galaxy-galaxy and galaxy-lensing response coefficients is not included in the signal, following \citetalias{ISTF2020}. Once computed in this way, the response coefficient can be projected in harmonic space using Eq.~\eqref{eq: 2dresponses}, and inserted in Eq.~\eqref{eq: covssc} to compute the SSC in the \citetalias{Lacasa_2019} approximation. The projected $R^{AB}_{ij}(\ell)$ functions are shown in Fig.~\ref{fig: responsecoefficients}.
%the galaxy and the galaxy-lensing response coefficient have a lower amplitude than $R^{\rm mm}(\ell)$, because the first-
%
%

%%%%%%%%%%%%%%%%%%%%%%%%%%%%%%%%%%%%%%%%%%%%%%%%%%%%%%%%%%%%%%%%%%%%%%%

\section{Forecasts specifics} \label{sec: specifics}

In order to forecast the uncertainties in the measurement of the cosmological parameters, we follow the prescriptions of the \Euclid forecast validation study (\citetalias{ISTF2020}), with some updates to the most recent results from the EC, which are used for the third Science Performance Verification (SPV) of \Euclid before launch. In particular, the update concerns the fiducial value of the linear bias, the redshift distribution $n(z)$ and the multipole binning.

Once again, the observable under study is the angular PS of probe $A$ in redshift bin $i$ and probe $B$ in redshift bin $j$, given in the Limber approximation by Eq.~\eqref{eq: CijSSC}. The $P_{AB} ( k_{\ell}, z )$ multi-probe power spectra are given in Eq.~\eqref{eq: powspec}; in the following, we refer interchangeably to the probes  (WL, XC, GCph) and their auto- and cross-spectra (respectively, LL, GL, GG). 

\subsection{Redshift distribution}\label{sec: z_distribution}

First, we assume that the same galaxy population is used to probe both the WL and the GCph PS. We therefore set 
\begin{equation}
n_{i}^{\rm L}(z) = n_{i}^{\rm G}(z) = n_i(z) \; ,
\label{eq: samenz}
\end{equation}
where $n_i^{\rm L}(z)$ and $n_i^{\rm G}(z)$ are respectively the distribution of sources and lenses in the $i$-th redshift bin. Then, the same equality applies for the total source and lens number density, $\bar{n}^{\rm L}$ and $\bar{n}^{\rm G}$.

A more realistic galaxy redshift distribution than the analytical one presented in \citetalias{ISTF2020} can be obtained from simulations. We use the results from \citet{Pocino2021}, in which the $n(z)$ is constructed from photometric redshift estimates in a 400 ${\rm deg}^2$ patch of the Flagship 1 simulation \citep{Potter2016}, using the training-based directional neighbourhood fitting (DNF) algorithm \citep[][]{DeVicente2016}.

The training set is a random subsample of objects with true (spectroscopic) redshifts known from the Flagship simulation. We choose the fiducial case presented in \citet{Pocino2021}, which takes into account a drop in completeness of the spectroscopic training sample with increasing magnitude. A cut in magnitude $\IE < 24.5$, isotropic and equal for all photometric bands, is applied, corresponding to the optimistic \Euclid setting. The DNF algorithm then produces a first estimate of the photo-$z$, $z_{\rm mean}$, using as a metric the objects' closeness in colour and magnitude space to the training samples. A second estimate of the redshift, $z_{\rm mc}$, is computed from a Monte Carlo draw from the nearest neighbour in the DNF metric. The final distributions for the different redshift bins, $n_i(z)$, are obtained by assigning the sources to the respective bins using their $z_{\rm mean}$, and then taking the histogram of the $z_{\rm mc}$ values in each of the bins -- following what has been done in real surveys such as the Dark Energy Survey \citep{Crocce2017, Hoyle2017}.

As a reference setting, we choose to bin the galaxy distribution into ${\cal N}_{\rm b} = $ 10 equipopulated redshift bins, with edges
% from /Users/davide/Documents/Lavoro/Programmi/common_data/vincenzo/SPV3_07_2022/Flagship_1_restored/InputNz/Sources/Flagship/ngbTab-EP10.dat
\begin{align}
    z_{\rm edges} = \{ & 0.001, 0.301, 0.471, 0.608, 0.731, 0.851, \nonumber \\ 
    & 0.980, 1.131, 1.335, 1.667,
       2.501 \} \; .
    \label{eq:zbins}
\end{align}
The total galaxy number density is  $\bar{n} = 28.73 \, {\rm arcmin}^{-2}$. As a comparison, this was set to $30 \, {\rm arcmin}^{-2}$ in \citetalias{ISTF2020}. We note that this choice of redshift binning will be discussed and varied in Sect.~\ref{sect: red_bin}.

\subsection{Weight functions}
We model the radial kernels, or weight functions, for WL and GCph following once again \citetalias{ISTF2020}. Adopting the eNLA ({extended} non-linear alignment) prescription for modelling the intrinsic alignment (IA) contribution, the weight function ${\cal W}_i^A(z)$ for the lensing part is given by \citep[see e.g.][]{Kitching2017, Kilbinger2017, Taylor2018}
\begin{equation}
{\cal{W}}_{i}^{\rm L}(z) = {\cal{W}}_{i}^{\gamma}(z) - 
\frac{{\cal{A}}_{\rm IA} {\cal{C}}_{\rm IA} \Omega_{{\rm m},0} {\cal{F}}_{\rm IA}(z)}
{D(z)} {\cal{W}}^{\rm IA}(z) \; ,
\label{eq: wildef}
\end{equation}
where we define\footnote{Equation~\eqref{eq: wigammadef} \dav{assumes general relativity and a spatially flat Universe}. For the general case, one must replace the term in brackets with $f_{K}(r^{\prime} - r)/f_K(r^{\prime})$, with $f_K(r)$ the function giving the comoving angular-diameter distance in a non-flat universe.}
\begin{equation}
{\cal{W}}_{i}^{\gamma}(z) = \frac{3}{2} \left ( \frac{H_0}{c} \right )^2 \Omega_{{\rm m},0} (1 + z) r(z) 
 \int_{z}^{z_{\rm max}}{\frac{n_{i}(z^\prime)}{\bar{n}} \left [ 1 - \frac{r(z)}{r(z^{\prime})} \right ] \; \diff z^{\prime}} ,
\label{eq: wigammadef}
\end{equation}
and
\begin{equation}
{\cal{W}}_{i}^\mathrm{IA}(z) =  \frac{1}{c}
\frac{n_i(z)}{\bar n} H(z) \; .
\label{eq: wiiadef}
\end{equation}
Finally, in Eq.~\eqref{eq: wildef}, ${\cal{A}}_{\rm IA}$ is the overall IA amplitude, ${\cal{C}}_{\rm IA}$ a constant, ${\cal{F}}_{\rm IA}(z)$ a function modulating the dependence on redshift, and $D(z)$ is the linear growth factor. More details on the IA modelling are given in Sect.~\ref{sec: ia}.

The GCph weight function is equal to the IA one, as long as Eq.~\eqref{eq: samenz} holds:
\begin{equation}
{\cal{W}}_{i}^{\rm G}(z) = {\cal{W}}_{i}^\mathrm{IA}(z) = \frac{1}{c} 
\frac{n_{i}(z)}{\bar n} H(z) \; .
\label{eq: wigdef}
\end{equation}

\begin{figure*}[!ht]
\centering
\includegraphics[width=\linewidth]{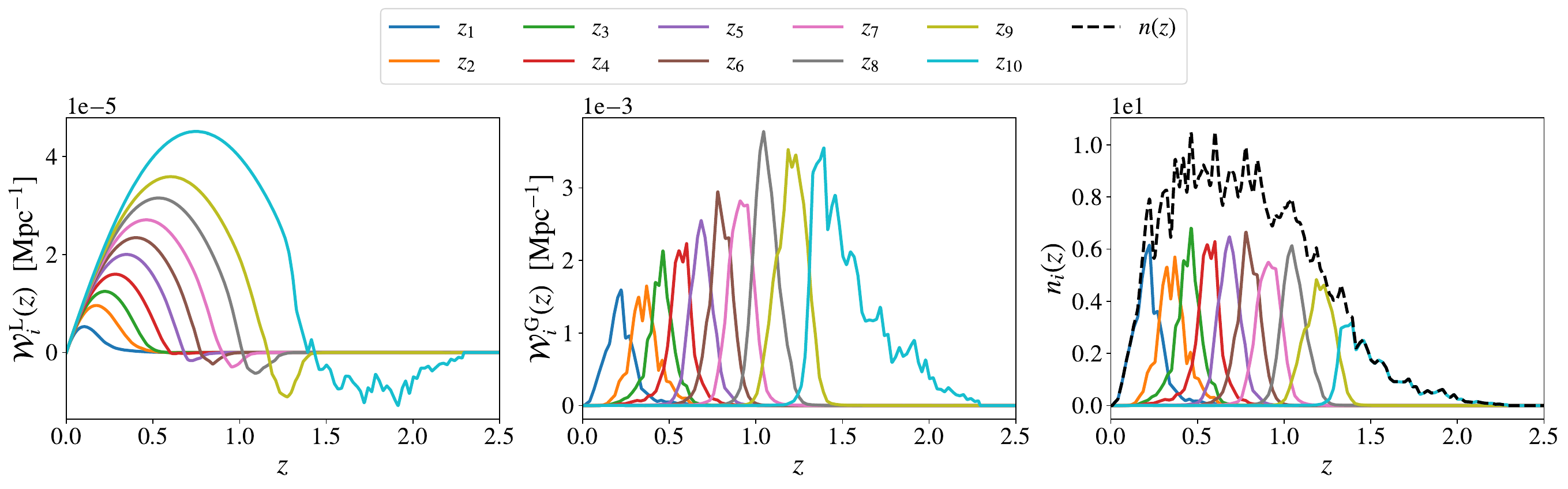}
\caption{\davthree{Kernels and galaxy distribution considered in this work. The first two plots show the kernels, or weight for the two photometric probes}. The analytic expressions for these are, respectively, Eq.~\eqref{eq: wildef} (left, WL) and Eq.~\eqref{eq: wigdef} (right, GCph). At high redshifts, the IA term dominates over the shear term in the lensing kernels, making them negative. The rightmost plot shows the redshift distribution per redshift bin for the  sources (and lenses), \dav{as well as their sum}, obtained from the Flagship 1 simulation as described in Sect.~\ref{sec: z_distribution}.}
\label{fig: WFs}
\end{figure*}

Figure~\ref{fig: WFs} shows the redshift dependence of Eqs.~\eqref{eq: wildef} and \eqref{eq: wigdef}, for all redshift bins. We note that we choose to include the galaxy bias term $b_i(z)$ in the PS (see Eq.~\ref{eq: powspec}) rather than in the galaxy kernel, as opposed to what has been done in \citetalias{ISTF2020}. This is done to compute the galaxy response as described in Sect.~\ref{sec: probe_response}. However, as the galaxy bias is assumed constant in each bin, the question is of no practical relevance when computing the $S_{ijkl}$ matrix, since the constant bias cancels out.

We note that the above definitions of the lensing and galaxy kernels (${\cal W}_i^A(z), \ A = {\rm L, G}$) differ from the ones used in \citetalias{Lacasa_2019}. This is simply because of a different definition of the $C_{ij}^{AB}(\ell)$ Limber integral, which is performed in $\diff V$ in \citetalias{Lacasa_2019} and in $\diff z$ in \citetalias{ISTF2020}. The mapping between the two conventions is simply given by the expression for the volume element:
\begin{equation}
    \diff V = r^2(z)\frac{\diff r}{\diff z}\diff z = c \, \frac{r^2(z)}{H(z)}\diff z \; ,
\end{equation}
and 
\begin{equation}\label{eq: wfmatch}
    W^A_i(z) = {\cal W}^A_i(z)/r^2(z) \; , 
\end{equation}
with $A = {\rm L, G}$. In Fig.~\ref{fig: WFs} we plot the values of ${\cal W}^A_i(z)$ to facilitate the comparison with \citetalias{ISTF2020}. As outlined in Appendix~\ref{sec: validation_appendix}, when computing the $S_{ijkl}$ matrix through \texttt{PySSC}, the user can either pass the kernels in the form used in \citetalias{Lacasa_2019} or the one used in \citetalias{ISTF2020} -- specifying a non-default \texttt{convention} parameter.

\subsection{Gaussian covariance}

The Gaussian part of the covariance is given by the following expression:
\begin{align}\label{eq: covgauss}
{\rm Cov_{\rm G}} & \left[\hat{C}_{ij}^{AB}(\ell), \hat{C}_{k l}^{C D}(\ell^{\prime})\right]=
\left[(2\,\ell + 1)\,f_{\rm sky} \, \Delta \ell \right]^{-1}
\delta_{\ell \ell^{\prime}}^{\rm K}  \nonumber \\
 & \times \Bigg\{ \left[C_{ik}^{AC}(\ell) + {N}_{ik}^{AC}(\ell)\right]
\left[C_{jl}^{BD}(\ell^\prime) + N_{jl}^{BD}(\ell^\prime) \right]  \nonumber \\
& + \left[C_{il}^{AD}(\ell) + N_{il}^{AD}(\ell) \right]
\left[C_{jk}^{BC}(\ell^\prime) + N_{jk}^{BC}(\ell^\prime) \right] 
\Bigg\}  \; ,
\end{align}
where we use a hat to distinguish the estimators from the true spectra. The noise PS $ N_{ij}^{AB}(\ell)$ are, for the different probe combinations, 
\begin{equation}
N_{ij}^{AB}(\ell) = \left \{
\begin{array}{ll}
\displaystyle{(\sigma_{\epsilon}^2/\dav{2}\bar{n}_{i}^{\rm L}) \, \delta_{ij}^{\rm K}} & \displaystyle{A = B = {\rm L} \;\; ({\rm WL})} \\
 & \\
\displaystyle{0} & \displaystyle{A \neq B} \\
 & \\
\displaystyle{(1/\bar{n}_{i}^{\rm G}) \, \delta_{ij}^{\rm K}} & \displaystyle{A = B = {\rm G} \;\; ({\rm GCph}}) \; . \\
\end{array}
\right . 
\label{eq: noiseps}
\end{equation}
In the above equations, $\delta_{ij}^{\rm K}$ is the Kronecker delta and $\sigma_{\epsilon}^2$ the variance of the total intrinsic ellipticity dispersion of WL sources, where $\sigma_{\epsilon} = \sqrt{2}\sigma_\epsilon^{(i)}$, with $\sigma_\epsilon^{(i)}$ being the ellipticity dispersion per component of the galaxy ellipse. 
\dav{Some care is needed when defining the shear noise spectrum: the above equation can then also be written as $N_{ij}^{\rm LL}(\ell)=\left[ (\sigma_{\epsilon}^{(i)})^2/\bar{n}_{i}^{\rm L}\right] \, \delta_{ij}^{\rm K} \,$, that is, using the ellipticity dispersion {per component} instead of the total one, which is the appropriate choice for harmonic-space analyses \citep{Hu2004, Joachimi2010}.}
We note that the average densities used in Eq.~\eqref{eq: noiseps} are not the total number densities, but rather those in the $i$-th redshift bin. In the case of ${\cal N}_{\rm b}$ equipopulated redshift bins, they can be simply written as $\bar{n}_{i}^{A} = \bar{n}^{A}/{\cal N}_{\rm b}$ for both $A = ({\rm L, G})$. Finally, we recall that $f_{\rm sky}$ is the fraction of the total sky area covered by the survey, while $\Delta \ell$ is the width of the multipole bin centred on a given $\ell$. From Sect.~\ref{sec: z_distribution} we have that $\bar{n} = 28.73 \, {\rm arcmin}^{-2}$, while we set $\sigma_\epsilon = 0.37$ \citep[from the value $\sigma_\epsilon^{(i)}=0.26$ reported in][]{Martinet_2019} and $f_{\rm sky} = 0.356$ (corresponding to $\Omega_\mathrm{S} = 14~700$ deg$^2$). We have now all the relevant formulae for the estimate of the Gaussian and the SSC terms of the covariance matrix. To ease the computation of Eq.~\eqref{eq: covgauss} we have prepared an optimised \texttt{Python} module, \texttt{Spaceborne\_covg}\footnote{\texttt{\url{https://github.com/davidesciotti/Spaceborne\_covg}}}, available as a public repository.

In the context of the present work, we do not consider the other non-Gaussian contribution to the total covariance matrix, the so-called connected non-Gaussian (cNG) term. This additional non-Gaussian term has been shown to be subdominant with respect to the Gaussian and SSC terms for WL both in \citet{Barreira2018cosmic_shear} and in \citet{upham2021}. For what concerns galaxy clustering, \citet{wadekar_20} showed that the cNG term was subdominant, but this was for a spectroscopic sample so (i) they had a much larger contribution from shot-noise-related terms compared to what is considered here for the \Euclid photometric sample, and (ii) they considered larger and more linear scales than in the present study. \citet{lacasa_20} showed that the cNG term in the covariance matrix of GCph only impacts the spectral index $n_\mathrm{s}$ and HOD parameters, but there are a few differences between that analysis and the present work, such as the modelling of galaxy bias. Thus it is still unclear whether the cNG term has a strong impact on cosmological constraints obtained with GCph. Quantifying the impact of this term for the 3$\times$2pt analysis with \Euclid settings is left for future work.

%As for the choice of the ingredients entering the covariance terms, we mostly adopt the same specifics as in \citetalias{ISTF2020}, which the reader is referred to for details. We nevertheless briefly descirbe these below to make the present work self-contained.

%
%
\subsection{Cosmological model and matter power spectrum}
We adopt a flat $ w_0w_a$CDM model, that is, we model the dark energy equation of state with a Chevallier--Polarski--Linder (CPL) parametrisation \citep{Chevallier2001, Linder2005}:
\begin{equation}
w(z) = w_0 + w_a \, z/(1 + z)  \; .
\label{eq: cpleos}
\end{equation}
We also include a contribution from massive neutrinos with total mass equal to the minimum allowed by oscillation experiments \citep{Esteban_20} $\sum{m_{\nu}} = 0.06 \ {\rm eV}$, which we do not vary in the FM analysis. The vector of cosmological parameters is then 
\begin{equation}\label{param_vector}
\vec{\theta}_{\rm cosmo} = \left\{\Omega_{{\rm m},0}, \Omega_{{\rm b},0}, w_0, w_a, h, n_{\rm s}, \sigma_8\right\} \; ,
\end{equation}
with $\Omega_{{\rm m},0}$ and $\Omega_{{\rm b},0}$ being respectively the reduced density of total and baryonic matter today, $h$ is the dimensionless Hubble parameter defined as $H_0 = 100 \, h\ \mathrm{km}\ \mathrm{s}^{-1}\ \mathrm{Mpc}^{-1} $ where $H_0$ is the value of the Hubble parameter today, $n_{\rm s}$ the spectral index of the primordial power spectrum and $\sigma_8$ the root mean square of the linear matter density field smoothed with a sphere of radius 8 \si{\hMpc}. \dav{We follow \citetalias{ISTF2020} for their fiducial values that are}
\begin{equation}\label{theta_fid}
\vec{\theta}_{\rm cosmo}^{\, \rm fid} = \left\{0.32, 0.05, -1.0, 0.0, 0.67, 0.96, 0.816\right\} \; . 
\end{equation}
\dav{This parameter vector then is used as input for the evaluation of the fiducial linear and non-linear matter PS; for the purpose of validating our forecasts against the \citetalias{ISTF2020} results, we use the \texttt{TakaBird} recipe, that is, the \texttt{HaloFit} version updated in \citet{Takahashi2012} with the \citet{Bird2012} correction for massive neutrinos. For the results shown in this paper, however, we update the non-linear model to the more recent \texttt{HMCode2020} recipe, \citep{Mead2020}, which includes a baryonic correction parameterised by the \logT\, parameter, characterising the feedback from active galactic nuclei (AGN). This is implemented in \texttt{CAMB}\footnote{\texttt{\url{https://camb.info/}}} \citep{Lewis2011} and, at the time of writing, is planned to be included in \texttt{CLASS}\footnote{\texttt{\url{https://lesgourg.github.io/class\_public/class.html}}} \citep{Blas2011} as well. Because of this, we add a further free parameter in the analysis, \logT\,, with a fiducial value of 7.75.}
\subsection{Intrinsic alignment model}\label{sec: ia}
We use the eNLA model as in \citetalias{ISTF2020}, setting ${\cal{C}}_{\rm IA} = 0.0134$ and
\begin{equation}
{\cal{F}}_{\rm IA}(z) = (1 + z)^{\, \eta_{\rm IA}} \left[\langle L \rangle(z)/L_{\star}(z)\right]^{\, \beta_{\rm IA}} \; ,
\label{eq: fiadef}
\end{equation}
where $\langle L \rangle(z)/L_{\star}(z)$ is the redshift-dependent ratio of the mean luminosity over the characteristic luminosity of WL sources as estimated from an average luminosity function \citep[see e.g.][and references therein]{Joachimi_2015}. The IA nuisance parameters vector is 
\begin{equation}
\vec{\theta}_{\rm IA} = \{{\cal{A}}_{\rm IA}, \eta_{\rm IA}, \beta_{\rm IA}\} \; ,
\end{equation}
with fiducial values -- \dav{following \citetalias{ISTF2020}}
\begin{equation}
\vec{\theta}_{\rm IA}^{\,\rm fid} =\{1.72, -0.41, 2.17\} \; . 
\end{equation}
All of the IA parameters except for ${\cal{C}}_{\rm IA}$ are varied in the analysis.

\subsection{Linear galaxy bias and multiplicative shear bias}\label{sec: bias}

Following \citetalias{ISTF2020} we model the galaxy bias as scale-independent. \dav{As for the redshift dependence}, we move beyond the simple analytical prescription of \citetalias{ISTF2020} and use the fitting function presented in \citet{Pocino2021}, obtained from direct measurements from the \Euclid Flagship galaxy catalogue, based in turn on the Flagship 1 simulation:
\begin{equation}\label{eq: pocinoBias}
    b(z) = \frac{Az^B}{1+z} + C \; ,
\end{equation}
setting $(A, B, C) = (0.81, 2.80, 1.02)$. 

The galaxy bias is modelled to be constant in each bin with the fiducial value obtained by evaluating Eq.~\eqref{eq: pocinoBias} at effective values $z^{\rm eff}_i$ computed as the median of the redshift distribution considering only the part of the distribution at least larger than 10\% of its maximum. \dav{We choose to use the median instead of the mean since, for equipopulated bins -- as can be seen from the rightmost panel of Fig.~\ref{fig: WFs} -- the galaxy distribution of the last bins is highly skewed, and the value of the bias computed at the mean is potentially less accurate; this choice does not, on the other hand, affect the galaxy bias values in the first bins sensibly. \\
The $z^{\rm eff}_i$ values obtained in this way are}
\dav{ 
\begin{align}
% old
%    z^{\rm eff} = \{ & 0.233, 0.373, 0.455, 0.571, 0.686,
%    \nonumber \\ 
%    & 0.796, 0.913, 1.070, 1.195, 1.628 \} \; .
% new (June 2024)
   z^{\rm eff} = \{ & 0.212, 0.363, 0.447, 0.566, 0.682,
    \nonumber \\ 
    & 0.793, 0.910, 1.068, 1.194, 1.628 \} \; .
    \label{eq:zeff} 
\end{align}
}
We therefore have ${\cal N}_{\rm b}$ additional nuisance parameters:
\begin{equation}
    \vec{\theta}_{\rm gal. \, bias} = \{b_1, b_2, \ldots, b_{{\cal N}_{\rm b}}\} \; ,
\end{equation}
with fiducial values
\begin{align}
    \vec{\theta}^{\rm fid}_{\rm gal. \, bias} = \{ & 1.031, 1.057, 1.081, 1.128, 1.187, \\ \nonumber
    & 1.258, 1.348, 1.493, 1.628,
       2.227 \} \; .
    \label{eq:zeff} 
\end{align}
\dav{The modelling of galaxy bias just described is the same used in \citetalias{ISTF2020}, with different fiducial values.\\}
We can take a further step forward towards the real data analysis by including the multiplicative shear bias parameters, $m$, defined as the multiplicative coefficient of the linear bias expansion of the shear field $\vec{\gamma}$ (see e.g. \citealt{Cragg2023}):
\begin{equation}\label{eq:shear_bias_def}
\hat{\vec{\gamma}} = (1+m) \, \vec{\gamma} + c
,\end{equation}
with $\hat{\vec{\gamma}}$ the measured shear field, $\vec{\gamma}$ the true one, $m$ the multiplicative and $c$ the additive shear bias parameters (we do not consider the latter in the present analysis, \dav{as we assume it will be corrected in the shear data processing pipeline}). The multiplicative shear bias can come from astrophysical or instrumental systematics (such as the effect of the point spread function -- PSF), which affect the measurement of galaxy shapes.
We take the $m_i$ parameters (one for each redshift bin) as constant and with a fiducial value of 0 in all bins. To include this further nuisance parameter, one just has to update the different angular PS as 
\begin{equation}
\left \{
\begin{array}{l}
\displaystyle{C_{ij}^{\rm LL}(\ell) \rightarrow (1 + m_i) (1 + m_j) C_{ij}^{\rm LL}(\ell)} \\
 \\ 
\displaystyle{C_{ij}^{\rm GL}(\ell) \rightarrow (1 + m_j) C_{ij}^{\rm GL}(\ell)} \\
 \\
\displaystyle{C_{ij}^{\rm GG}(\ell) \rightarrow C_{ij}^{\rm GG}(\ell)} \; , \\
\end{array}
\right .
\label{eq: cijbias}
\end{equation}
where $m_i$ is the $i$-th bin multiplicative bias, and the GCph spectrum is unchanged since it does not include any shear term. We then have\begin{equation}
\vec{\theta}_{\rm shear \, bias} = \{m_1, m_2, \ldots, m_{{\cal N}_{\rm b}}\} \; ,
\end{equation}
with fiducial values
\begin{equation}\label{eq: bzvalue}
\vec{\theta}_{\rm shear \, bias}^{\,\rm fid} = \left \{ 0, 0, \ldots, 0
\right \} \; .
\end{equation}

\dav{Finally, we introduce the $\Delta z_i$ parameters to allow for uncertainties over the first moments of the photometric redshift distribution (argued to have the largest impact on the final constraints in \citealt{Reischke2024_photoz})}. We then have \citep{Troxel2018_DES_Y1, Abbott2018_DES_Y1, Tutusaus2020}:
\begin{equation}
    \dav{n_i(z) \rightarrow n_i(z - \Delta z_i),}
\end{equation}
which adds new entries to our nuisance parameter vector:
\begin{equation}
\dav{\vec{\theta}_{{\rm photo}-z} = \{\Delta z_1, \Delta z_2, \ldots, \Delta z_{{\cal N}_{\rm b}}\}} \; ,
\end{equation}
with fiducial values:
\begin{equation}\label{eq: Delta_photoz_values}
\dav{\vec{\theta}_{{\rm photo}-z}^{\,\rm fid} = \left \{ 0, 0, \ldots, 0
\right \} }\; .
\end{equation}

These nuisance parameters -- unless specified otherwise -- are varied in the Fisher analysis so that the final parameters vector is 
\begin{displaymath}
\vec{\theta} = \vec{\theta}_{\rm cosmo} \cup \vec{\theta}_{\rm IA} \cup \vec{\theta}_{\rm gal. \, bias} \cup \vec{\theta}_{\rm shear \, bias}\cup \dav{\vec{\theta}_{{\rm photo}-z}} \; ,
\end{displaymath}
and 
\begin{displaymath}
\vec{\theta}^{\,\rm fid} = \vec{\theta}_{\rm cosmo}^{\,\rm fid} \cup \vec{\theta}_{\rm IA}^{\,\rm fid} \cup \vec{\theta}_{\rm gal. \, bias}^{\,\rm fid} \cup \vec{\theta}_{\rm shear \, bias}^{\,\rm fid} \cup \dav{\vec{\theta}_{{\rm photo}-z}^{\,\rm fid}}\; ,
\end{displaymath}
both composed of ${\cal N}_{\rm p} = 7 + 3 + 3{\cal N}_{\rm b} = 3{\cal N}_{\rm b} + 10$ elements. 

\subsubsection{Higher-order bias}\label{sec: higher_order_bias}
To compute the galaxy--galaxy and galaxy--galaxy lensing probe response terms (Eqs.~\ref{eq: Rgg} and~\ref{eq: Rgm}) we need the second-order galaxy bias $b_{(2)}(z)$. To do this, we follow Appendix C of \citetalias{Lacasa_2019}, in which this is estimated following the halo model\footnote{We neglect the response of $\langle N\vert M \rangle$ to a perturbation $\delta_{\rm b}$ in the background density, as done in \citetalias{Lacasa_2019}.} as \citep{VB21,Alex2021}
\begin{equation}
b_{(i)}(z) = \int{\diff M \; \Phi_{\rm MF}(M, z) b_{(i)}^{\rm h}(M, z) \langle N \vert M \rangle/n_{\rm gal}(z)},
\label{eq: bicalc}
\end{equation}
with 
\begin{equation}
n_{\rm gal}(z) = \int{\diff M \; \Phi_{\rm MF}(M, z) \langle N|M \rangle },
\label{eq: ngalvszed}
\end{equation}
the galaxy number density, $\Phi_{\rm MF}(M, z)$ the halo mass function (HMF), $b_{(i)}^{\rm h}(M, z)$ the $i$-th order \textit{halo} bias, and $\langle N|M \rangle$ the average number of galaxies hosted by a halo of mass $M$ at redshift $z$ (given by the halo occupation distribution, HOD). These are integrated over the mass range $\log M \in [9,16]$, with the mass expressed in units of solar masses (we don't include $h$ in our units). The expression for the $i$-th order galaxy bias (Eq.~\ref{eq: bicalc}) is the same as Eq.~(C.2) of \citetalias{Lacasa_2019}, but here we are neglecting the scale dependence of the bias evaluating it at $k = 0$ so that $u(k \, | \, M = 0, z) = 1$, $u(k \, | \, M, z)$ being the Fourier Transform of the halo profile. Strictly speaking, this gives us the large-scale bias, but it is easy to check that the dependence on $k$ is negligible over the range of interest.

Although Eq.~\eqref{eq: bicalc} allows the computation of both the first and second-order galaxy bias, we prefer to use the values of $b_{(1)}(z)$ measured from the Flagship simulation for the selected galaxy sample; this is to maintain consistency with the choices presented at the beginning Sect.~\ref{sec: bias}. For each redshift bin, we vary (some of) the HOD parameters to fit the measured $b_{(1)}(z)$, thus getting a model for $b_{(1)}^{\rm h}(z)$. We then compute $b_{(2)}^{\rm h}(z)$ using as an additional ingredient the following relation between the first and second-order halo bias, which approximates the results from separate universe simulations \citep{Lazeyras_2016} within the fitting range $1 \lesssim b_{(1)}^{\rm h} \lesssim 10$: 
\begin{align} \label{eq: b2vsvb1}
b_{(2)}^{\rm h}(M, z) & = 0.412 - 2.143 \, b_{(1)}^{\rm h}(M, z) \nonumber \\
& + 0.929 \, \left[b_{(1)}^{\rm h}(M, z)\right]^2 
+ 0.008 \, \left[b_{(1)}^{\rm h}(M, z)\right]^3 \; . 
\end{align}
Finally, we plug the $b_{(2)}^{\rm h}$ values obtained in this way back into Eq.~\eqref{eq: bicalc} to get the second-order galaxy bias. The details of the HMF and HOD used and of the fitting procedure are given in Appendix \ref{sec: halomodel}.

\subsection{Data vectors and Fisher matrix}
\label{sec:datavectors}

Up to now, we have outlined a fully general approach, without making any assumptions about the data. We now need to set data-related quantities.

First, we assume that we will measure $C_{ij}^{AB}(\ell)$ in ten equally populated redshift bins over the redshift range $(0.001, 2.5)$. When integrating Eq.~\eqref{eq: CijSSC} in $\diff z$, $z_{\rm max}$ must be larger than the upper limit of the last redshift bin to account for the broadening of the bin redshift distribution due to photo-$z$ uncertainties. We have found that the $C_{ij}^{AB}(\ell)$ stop varying for $z_{\rm max} \ge 4$, which is what we take as the upper limit in the integrals over $z$.
This also means that we need to extrapolate the bias beyond the upper limit of the last redshift bin; we then take its value as constant and equal to the one in the last redshift bin, that is, $b(z > 2.501) = b_{10}$.

Second, we assume the same multipole limits as in \citetalias{ISTF2020}, and therefore examine two scenarios, as follows:
\begin{itemize}
\item[-] pessimistic:
\begin{displaymath}
(\ell_{\rm min}, \ell_{\rm max}) = \left \{
\begin{array}{ll}
(10, 1500) & {\rm for \ WL} \\
 & \\
(10, 750) & {\rm for \ GCph \ and \ XC}
\end{array}
\right . \; ,
\end{displaymath}
\item[-] optimistic:
\begin{displaymath}
(\ell_{\rm min}, \ell_{\rm max}) = \left \{
\begin{array}{ll}
(10, 5000) & {\rm for \ WL} \\
 & \\
(10, 3000) & {\rm for \ GCph \ and \ XC}
\end{array}
\right . \; .
\end{displaymath}
\end{itemize}
Then, for the multipole binning, instead of dividing these ranges into ${\cal N}_{\ell}$ (logarithmically equispaced) bins in all cases as is done in \citetalias{ISTF2020}, we follow the most recent prescriptions of the EC and proceed as follows:
\begin{itemize}
    \item we fix the centres and edges of 32 bins (as opposed to 30) in the $\ell$ range $[10, 5000]$ following the procedure described in Appendix~\ref{sec: appendix_binning}. This will be the $\ell$ configuration of the optimistic WL case.
    \item The bins for the cases with $\ell_{\rm max} < 5000$, such as WL pessimistic, GCph, or XC, are obtained by cutting the bins of the optimistic WL case with $\ell_{\rm centre} > \ell_{\rm max}$. This means that instead of fixing the number of bins and having different bin centres and edges as done in \citetalias{ISTF2020}, we fix the bins' centres and edges and use a different number of bins, resulting in, for example, ${\cal N}_\ell^{\, \rm WL} > {\cal N}_\ell^{\, \rm GCph}$.
\end{itemize}
The number of multipole bins is then ${\cal N}_\ell^{\, \rm WL} = 26$ and ${\cal N}_\ell^{\, \rm GCph} = {\cal N}_\ell^{\, \rm XC} = 22$ in the pessimistic case and ${\cal N}_\ell^{\, \rm WL} = 32$ and ${\cal N}_\ell^{\, \rm GCph} = {\cal N}_\ell^{\, \rm XC} = 29$ in the optimistic case. In all these cases, the angular PS are computed at the centre of the $\ell$ bin, as done in \citetalias{ISTF2020}. \\
\dav{We note that, because of the width of the galaxy -- and, more importantly, lensing -- kernels, a given fixed $\ell_{\rm max}$ will not correspond to a unique $k_{\rm max}$ value. A more accurate approach could be, for example, to use the $k$-cut method presented in \cite{Taylor2018_kcut}, which leverages the BNT (Bernardeau-Nishimichi-Taruya) transform \citep{Bernardeau_2014} to make the lensing kernels separable in $z$, hence allowing for a cleaner separation of scales. We leave the investigation of this important open issue to dedicated work.}\\
\indent
As mentioned, we consider the different probes in isolation, as well as combine them in the `3$\times$2pt' analysis, which includes three 2-point angular correlation functions (in harmonic space): $C_{ij}^{\rm LL}(\ell), C_{ij}^{\rm GL}(\ell)$ and $C_{ij}^{\rm GG}(\ell)$. The $\ell$ binning for the 3$\times$2pt case is the same as for the GCph one.

The covariance matrix and the derivatives of the data vector w.r.t. the model parameters are the only elements needed to compute the FM elements. The one-dimensional data vector $\vec{C}$ is constructed by simply compressing the redshift and multipole indices (and, in the 3$\times$2pt case, the probe indices) into a single one, which we call $p$ (or $q$).
For Gaussian-distributed data with a parameter-independent covariance, the FM is given by:
\begin{equation}\label{eq: fishmat}
    F_{\alpha \beta} =     
    \frac{\partial \vec{C}}{\partial \theta_{\alpha}}
    \, {\rm Cov}^{-1} \,
    \frac{\partial \vec{C}}{\partial \theta_{\beta}}
    =  
    \sum_{pq}
    \frac{\partial C_p}{\partial \theta_{\alpha}}
    \, {\rm Cov}^{-1}_{pq} \,
    \frac{\partial C_q}{\partial \theta_{\beta}} \; .
\end{equation}
\dav{We refer the reader to \citetalias{ISTF2020} for details on the convergence and stability of the Fisher matrix and derivatives computations.}

We note that the size of the 3$\times$2pt covariance matrix quickly becomes large. For a standard setting with ${\cal N}_{\rm b} = 10$ redshift bins, there are respectively (55, 100, 55) independent redshift bin pairs for (WL, XC, GCph), to be multiplied by the different ${\cal N}_{\ell}$. In general, ${\rm Cov}$ will be a ${\cal N}_C \times {\cal N}_C$ matrix with 
\begin{align}
{\cal N}_C & = \bigg[
{\cal N}_{\rm b} ({\cal N}_{\rm b} + 1)/2 
\bigg]\bigg[
{\cal N}_\ell^{\, \rm WL} + {\cal N}_\ell^{\, \rm GCph}
\bigg] 
+ {\cal N}_{\rm b}^2  {\cal N}_\ell^{\, \rm XC} \nonumber \\
& = \bigg[{\cal N}_{\rm b} ({\cal N}_{\rm b} + 1) + {\cal N}_{\rm b}^2 \bigg]  {\cal N}_\ell^{\rm 3{\times}2pt} \, ,
\label{eq: sizecov}
\end{align}
\dav{where the second line is for the 3$\times$2pt case, which has the same number of $\ell$ bins for all probes}, and 
\begin{equation}
{\cal N}_C = \big[{\cal N}_{\rm b} ({\cal N}_{\rm b} + 1)/2 \big] \, {\cal N}_\ell^{\, \rm WL/GCph} \; ,
\end{equation}
for the WL and GCph cases. As an example, we have ${\cal N}_C^{\rm 3{\times}2pt, \, opt} = 6090$.

\begin{figure*}
    \centering
    \includegraphics[width=0.95\textwidth]{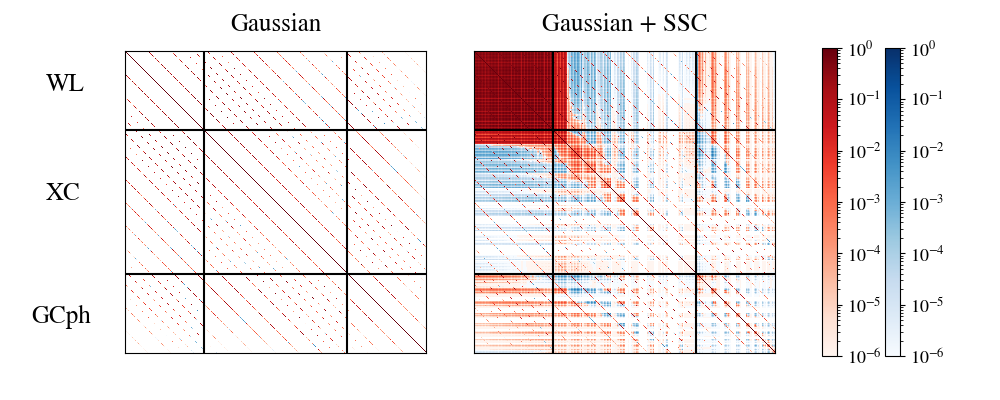}

    \caption{Correlation matrix in log scale for all the statistics of the 3$\times$2pt data-vector in the G and GS cases. The positive and negative elements are shown in red and blue, respectively. The Gaussian covariance is block diagonal (i.e. it is diagonal in the multipole indices, but not in the redshift ones; the different diagonals appearing in the plot correspond to the different redshift pair indices, for $\ell_1 = \ell_2$). The overlap in the WL kernels makes the WL block in the Gaussian + SSC covariance matrix much more dense than the GCph one.}
    \label{fig:cormat_GOvsGS}
\end{figure*}

Being diagonal in $\ell$, most elements of this matrix will be null in the Gaussian case. As shown in Fig.~\ref{fig:cormat_GOvsGS}, this is no longer true with the inclusion of the SSC contribution, which makes the matrix computation much more resource-intensive. The use of the \texttt{Numba JIT} compiler\footnote{\texttt{\url{https://numba.pydata.org}}} can dramatically reduce the CPU (for Central Processing Unit) time from about $260 \, {\rm s}$ to about $2.5 \, {\rm s}$ for the Gaussian + SSC 3$\times$2pt covariance matrix (the largest under study) on a normal laptop working in single-core mode.

Given the highly non-diagonal nature of the Gaussian + SSC covariance, we might wonder whether the inversion of this matrix (which is needed to obtain the FM, see Eq.\ref{eq: fishmat}) is stable. To investigate this, we compute the condition number of the covariance, which is defined as the ratio between its largest and smallest eigenvalues and in this case of order $10^{13}$. This condition number, multiplied by the standard \texttt{numpy float64} resolution ($2.22\times10^{-16}$), gives us the minimum precision that we have on the inversion of the matrix, of about $10^{-3}$. This means that numerical noise in the matrix inversion can cause, at most, errors of order $10^{-3}$ on the inverse matrix. Hence, we consider the inversion to be stable for the purpose of this work.

%
%%%%%%%%%%%%%%%%%%%%%%%%%%%%%%%%%%%%%%%%%%%%%%%%%%%%%%%%%%%%%%%%%%%%%%%
%
\section{Forecast code validation}
\label{sec: validation}

In order to validate the SSC computation with \texttt{PySSC}, we compare the $1\sigma$ forecast uncertainties (which correspond to a 68.3\% probability, due to the assumptions of the FM analysis) obtained using two different codes independently developed by two groups, which we call A and B. To produce the FM and the elements needed for its computation (the observables, their derivatives and the covariance matrix), group A uses a private\footnote{Available upon request to the author, Davide Sciotti} code fully written in \texttt{Python} and group B uses $\texttt{CosmoSIS}$\footnote{\texttt{\url{https://bitbucket.org/joezuntz/cosmosis/wiki/Home}}} \citep{Zuntz_2015}.  As stated in the introduction, the only shared feature of the two pipelines is the use of \texttt{PySSC} (to compute the $S_{ijkl}$ matrix). For this reason, and because the SSC is not considered in isolation but added to the Gaussian covariance, we compare the forecast results of the two groups both for the Gaussian and Gaussian + SSC cases.

Following \citetalias{ISTF2020}, we consider the results to be in agreement if the discrepancy of each group's results with respect to the median -- which in our case equals the mean --  is smaller than 10\%. This simply means that the A and B pipelines' outputs are considered validated against each other if
\begin{equation} \label{eq: sigmas}
    \abs{\frac{\sigma_\alpha^i}{\sigma^m_\alpha}-1} < 0.1 \quad {\rm for} \quad i = {\rm A,B}; \quad \sigma^m_\alpha = \frac{\sigma_\alpha^A+\sigma_\alpha^B}{2} \; ,
\end{equation}
with $\sigma_\alpha^A$ the $1\sigma$ uncertainty on the parameter $\alpha$ for group A. The above discrepancies are equal and opposite in sign for A and B.

The {marginalised} uncertainties are extracted from the FM $F_{\alpha\beta}$, which is the inverse of the covariance matrix ${\rm C}_{\alpha\beta}$ of the parameters: $(F^{-1})_{\alpha\beta} = {\rm C}_{\alpha\beta}$. The unmarginalised, or \textit{conditional}, uncertainties are instead given by $\sigma_\alpha^{\rm unmarg.} = \sqrt{1/F_{\alpha\alpha}}$. We then have
\begin{equation} \label{eq: cramer_rao}
    \sigma_\alpha = \sigma_\alpha^{\rm marg.} =  \sqrt{(F^{-1})_{\alpha\alpha}} \; .
\end{equation}
The uncertainties found in the FM formalism constitute lower bounds, or optimistic estimates, on the actual parameters' uncertainties, as stated by the Cramér-Rao inequality.

In the following, we normalise $\sigma_\alpha$ by the fiducial value of the parameter $\theta_\alpha$, in order to work with relative uncertainties: $\bar{\sigma}^i_\alpha = \sigma_\alpha^i/\theta_\alpha^{\,\rm fid}; \ \bar{\sigma}^m_\alpha = \sigma^m_\alpha/\theta_\alpha^{\,\rm fid}$, again with $i = {\rm A, B}$. If a given parameter has a fiducial value of 0, such as $w_a$, we simply take the absolute uncertainty. The different cases under examination are dubbed `G', or `Gaussian', and `GS', or `Gaussian + SSC'. The computation of the parameters constraints differs between these two cases only by the covariance matrix used in Eq.~\eqref{eq: fishmat} to compute the FM:
\begin{equation}
 \rm{Cov} = 
\begin{cases}
\rm Cov_{\rm G} & \rm{Gaussian}\\
\rm Cov_{\rm GS} = Cov_{\rm G} +  Cov_{SSC} & \rm{Gaussian+SSC} \; .
\end{cases}
\end{equation}

As mentioned above, we repeat the analysis for both \Euclid's photometric probes taken individually, WL and GCph, as well as for the combination of WL, GCph, and their cross-correlation XC, the 3$\times$2pt. 

\begin{figure}
\includegraphics[width=\hsize]{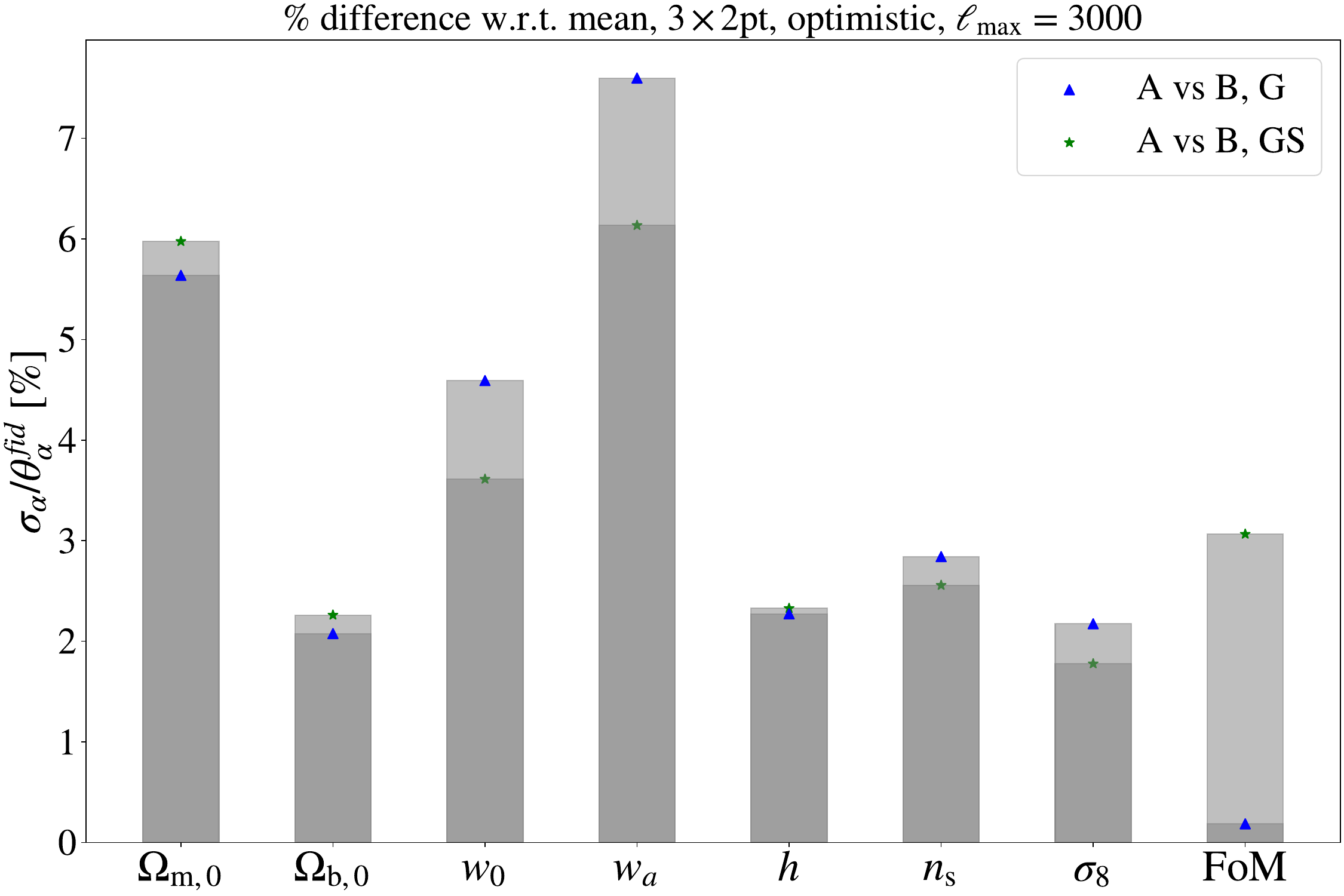}
\caption{Percentage discrepancy of the normalised $1\sigma$ uncertainties with respect to the mean for the WL probe, both in the G and GS cases (optimistic settings). The index $i = {\rm A, B}$ indicates the two pipelines, whilst $\alpha$ indexes the cosmological parameter. The desired agreement level is reached in all cases (WL, GCph probes and pessimistic case not shown).}
\label{fig: dav_vs_sylv}
\end{figure}

For the reader wanting to validate their own code, we describe the validation process in Appendix~\ref{sec: validation_appendix}. Here we sketch the results of the code validation: in Fig.~\ref{fig: dav_vs_sylv}, we show the percent discrepancy as defined in Eq.~\eqref{eq: sigmas} for the 3$\times$2pt case. Similar results have been obtained for the GCph and WL cases, both for the optimistic and pessimistic settings specified in Sect.~\ref{sec:datavectors}. The constraints are all found to satisfy the required agreement level (less than $10\%$ discrepancy with respect to the mean). In light of these results, we consider the two forecasting pipelines validated against each other. All the results presented in this paper are the ones produced by group A.
%In light of these results, we consider the computation of the SSC through the approximations and code presented in \citetalias{Lacasa_2019} validated for \Euclid's primary photometric probes.

%This study can be extended to the case of a variable $R^{AB}(\ell)$, for which a linear approximation is given in Appendix C of \citetalias{Lacasa_2019}, and to a full (CPU-intensive) computation, mainly to validate the approximation at high $\ell$.

%%%%%%%%%%%%%%%%%%%%%%%%%%%%%%%%%%%%%%%%%%%%%%%%%%%%%%%%%%%%%%%%%%%%%%%

\section{SSC impact on forecasts}
\label{sec: impact}

We investigate here how the inclusion of SSC degrades the constraints with respect to the Gaussian case. To this end, we look in the following at the quantity
\begin{equation}
\mathcal{R}(\theta) = \sigma_{\rm GS}(\theta)/\sigma_{\rm G}(\theta) \; ,
\label{eq: ratiodef}
\end{equation}
where $\sigma_{\rm G}(\theta)$ and $\sigma_{\rm GS}(\theta)$ are the usual  marginalised uncertainties on the parameter $\theta$ computed, as detailed above, with Gaussian or Gaussian\,+\,SSC covariance matrix. We run $\theta$ over the set of cosmological parameters listed in Eq.~\eqref{param_vector}, that is, $\theta \in \{\Omega_{{\rm m},0}, \Omega_{{\rm b},0}, w_0, w_a, h, n_{\rm s}, \sigma_8\}$.

In addition, we examine the FoM as defined in \citet{albrecht2006}, a useful way to quantify the joint uncertainty on several parameters. In this work, we parameterise the FoM following \citetalias{ISTF2020}: for two given parameters $\theta_1$ and $\theta_2$, we have 
\begin{equation}
    {\rm FoM}_{\theta_1\theta_2} = \sqrt{ {\rm  det}(\tilde{F}_{\theta_1\theta_2})} \; .
    \label{eq: fom}
\end{equation}
\dav{This quantity is inversely proportional to the area of the 2$\sigma$ confidence ellipse in the plane spanned by the parameters $(\theta_1, \theta_2)$. $\tilde{F}_{\theta_1\theta_2}$ is the Fisher submatrix obtained by marginalising over all the parameters but $\theta_1$ and $\theta_2$, and is computed by inverting $F_{\alpha\beta}$ (that is, taking the parameters' covariance matrix), removing all the rows and columns but the ones corresponding to $\theta_1$ and $\theta_2$ and reinverting the resulting $2 \times 2$ matrix. 
In the following, we mainly focus on the joint uncertainty on the dark energy equation of state parameters $w_0$ and $w_a$, and unless specified otherwise we use the notation ${\rm FoM} = {\rm FoM}_{w0wa}$. However, the FoM can help quantify the joint uncertainty on different sets of parameters, such as $(\Omega_{{\rm m},0} - S_8)$, with $S_8 \equiv \sigma_8(\Omega_{{\rm m},0}/0.3)^{0.5}$ (see e.g. \citealt{Abbott2022_DES_Y3})}.\\
\indent
We also use the notation ${\cal{R}}({\rm FoM})$ as a shorthand for ${\rm FoM}_{\rm GS}/{\rm FoM}_{\rm G}$. We note that, since we expect the uncertainties to be larger for the GS case, we have ${\cal{R}}(\theta) > 1$, and the FoM being inversely proportional to the area of the uncertainty ellipse, ${\cal{R}}({\rm FoM}) < 1$. 

%In addition we examine the Figure of Merit, a useful way to quantify the joint uncertainty on two parameters, $\theta_\alpha$ and $\theta_\beta$: following the definition given in \citet{albrecht2006}, we have

\begin{table*}
\centering
\caption{\dav{Ratio between the GS and G conditional uncertainties for all cosmological parameters and probes in the reference case, for the optimistic settings.}}
\begin{tabular}{l | c c c c c c c c c}
\hline
\multicolumn{1}{l |}{$\mathcal{R}(x)$} & $\Omega_{{\rm m},0}$ & $\Omega_{{\rm b},0}$ & $w_0$ & $w_a$ & $h$ & $n_{\rm s}$ & $\sigma_8$ & $\logten(T_{\rm AGN}/{\rm K})$\\
\hline
\hline
%\multicolumn{1}{l |}{WL, Pessimistic} & 
% 2.756 & 1.727 & 1.804 & 1.258 & 1.359 & 2.335 & 2.472 & 2.282  \\
\multicolumn{1}{l |}{WL} & 
3.127 & 1.839 & 2.014 & 1.189 & 1.683 & 2.496 & 2.889 & 2.171  \\
\hline
%\multicolumn{1}{l |}{XC, Pessimistic} & 
%1.178 & 1.076 & 1.043 & 1.118 & 1.016 & 1.192 & 1.165 & 1.454  \\
\multicolumn{1}{l |}{XC} & 
1.361 & 1.278 & 1.183 & 1.314 & 1.207 & 1.293 & 1.284 & 1.398  \\
\hline
%\multicolumn{1}{l |}{GCph, Pessimistic} & 
% 1.001 & 1.004 & 1.007 & 1.005 & 1.001 & 1.013 & 1.013 & 1.032 \\
\multicolumn{1}{l |}{GCph} & 
1.096 & 1.158 & 1.057 & 1.043 & 1.089 & 1.099 & 1.077 & 1.293  \\
\hline      
%\multicolumn{1}{l |}{3$\times$2pt, Pessimistic} & 
% 1.372 & 1.002 & 1.084 & 1.045 & 1.000 & 1.007 & 1.008 & 1.014 \\
\multicolumn{1}{l |}{3$\times$2pt} & 
1.321 & 1.078 & 1.083 & 1.055 & 1.047 & 1.050 & 1.039 & 1.130  \\
\hline
\end{tabular}
\label{tab:ratio_ref_conditional}
\end{table*}

\subsection{Reference scenario}\label{sec:reference_scenario}
Let us start by considering the case with ${\cal N}_{\rm b} = 10$ equipopulated redshift bins. \dav{To isolate the impact of SSC and gain better physical insight, we begin by computing the conditional uncertainties as described in the last section. 
%In this way, the uncertainty on a given parameter is not increased by the imperfect knowledge of the others. Alternatively, this means we consider the cross-parameter covariance to be zero
}\\
Table~\ref{tab:ratio_ref_conditional} gives the values of the ${\cal{R}}$ ratios for the different parameters in the optimistic scenarios, for the single probes and their combination. 

% TODO cite upham in marginalised uncert section
In accordance with previous results in the literature, we find that the WL constraints are dramatically impacted by the inclusion of SSC:
\dav{as found in \cite{Barreira2018cosmic_shear} (cfr. their Fig.~2), all cosmological parameters are affected, with $w_a$ and $h$ being impacted the least and $(\Omega_{{\rm m,0}}, \sigma_8, \logT)$ the most. 
This is because the SSC effect is essentially an unknown shift, or perturbation, in the background density within the survey volume, and is hence degenerate with the parameters which more closely relate to the amplitude of the signal. Being this a non-linear effect, its impact is also tied to the amount of power on small scales, which in turn is influenced by the baryonic boost, parameterised by \logT\ in our model. } \\
%We note that the $n_{\rm s}$ constraint is also highly degraded by SSC, although a word of caution is in order in this case: the high constraining power found from the FM analysis, also found in \citetalias{ISTF2020}, is most likely a spurious effect, being WL largely insensitive to this parameter.}\\
\indent
The results in Table~\ref{tab:ratio_ref_conditional} also show that GCph is not as strongly affected by SSC \dav{-- with the exception of the \logT\ constraint. This is an expected result (see e.g. \citealt{Bayer2022}), mainly driven by the fact that} the GCph probe response coefficients are lower (in absolute value) than the WL ones, as can be seen in Fig.~\ref{fig: responsecoefficients}. This is due to the additional terms that account for the response of the galaxy number density $n_{\rm g}$ (see Eq.~\ref{eq: Rgg}), which is itself affected by the super-survey modes.
Additionally, as can be seen from Fig.~\ref{fig: WFs}, all WL kernels have non-zero values for $z \rightarrow 0$, contrary to the GCph ones. In this limit, the effective volume probed by the survey tends to 0, hence making the variance of the background modes $\sigma^2$ tend to infinity. We thus have a larger $S_{ijkl}$ matrix, which is one of the main factors driving the amplitude of the SSC. 
\davtwo{We also note that the importance of baryonic feedback for GCph depends on the galaxy bias model used; it will likely be reduced when using a non-linear bias expansion, needed for an accurate analysis on small scales \citep{Desjacques_2018}. Both topics are still active areas of research, and we leave the characterisation of the SSC impact with the inclusion of higher-order bias terms for future work.}\\
\indent
\dav{The impact for the full 3$\times$2pt case sits in principle in between the two extremes as a consequence of the data vector containing the strongly affected WL probe, and the less affected GCph one. However, this is clearly apparent only in the case of $\Omega_{{\rm m},0}$, since it is the only cosmological parameter for which the WL constraining power is higher than the GCph one; for the other parameters, the trend resembles very closely the one found for GCph, because of its dominant contribution to the 3$\times$2pt precision.\\
Lastly, the contribution from the XC probe is again an intermediate case, as can be anticipated by looking at its response coefficient in Fig.~\ref{fig: responsecoefficients}, so the final impact on the FM elements will be intermediate between the WL and GCph cases, as the ${\cal{R}}(\theta)$ values in Table~\ref{tab:ratio_ref_conditional} indeed show.}

\begin{figure}
%\centering
    \includegraphics[width=0.47\textwidth]{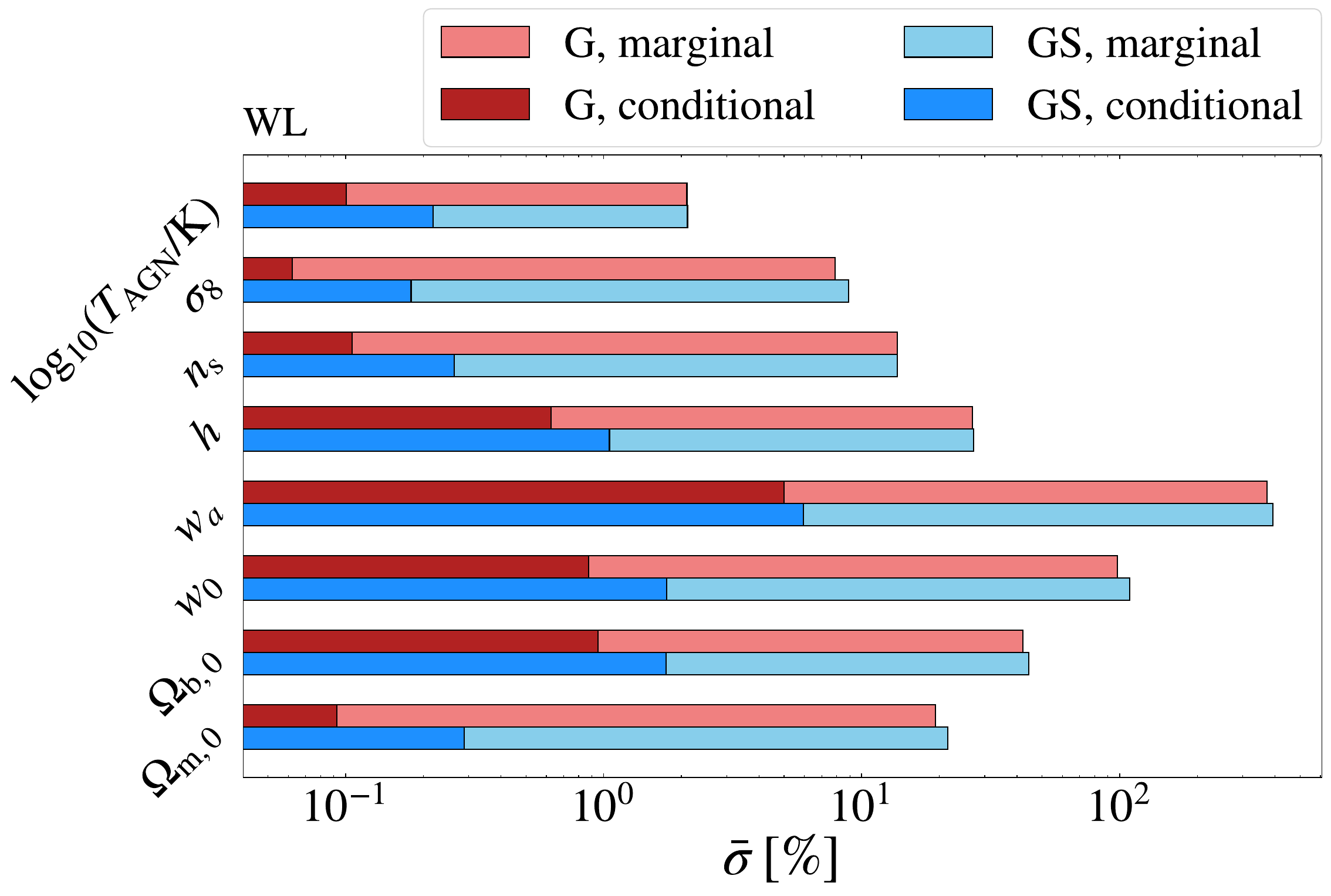}
    \includegraphics[width=0.47\textwidth]{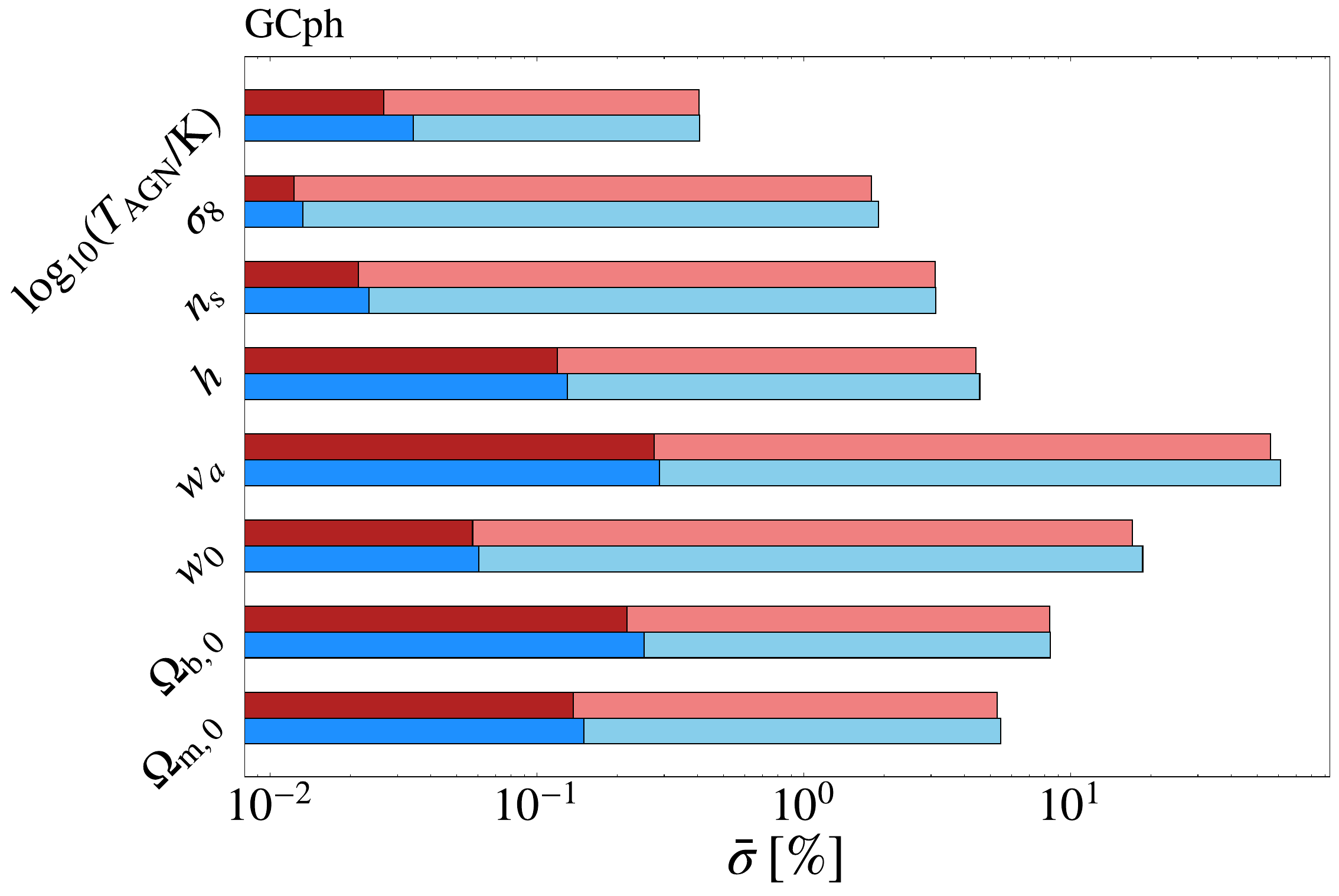}
    \includegraphics[width=0.47\textwidth]{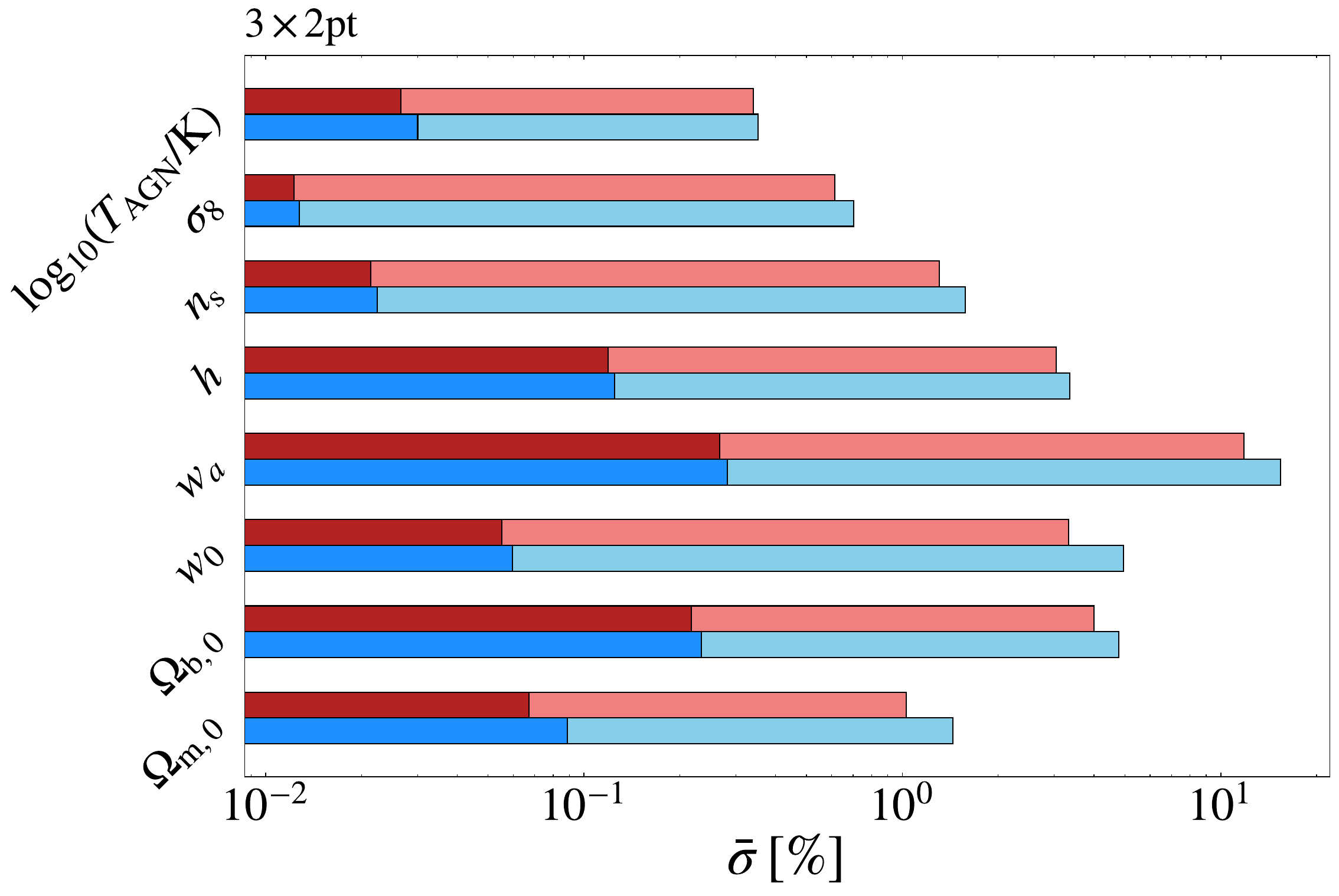}
    \caption{\dav{Marginalised and conditional optimistic 1$\sigma$ uncertainties on the cosmological parameters, relative to their corresponding fiducial values (in percent units), in both the G and GS cases for WL, GCph, and the 3$\times$2pt. We highlight the logarithmic scale on the $x$-axis, made necessary by the large range of values. The values in these plots have been used to compute the ratios in Tables~\ref{tab:ratio_ref_conditional} and \ref{tab:ratio_ref_marginal}.}}
    \label{fig: barplot}
\end{figure}
\begin{figure}
    \includegraphics[width=0.47\textwidth]{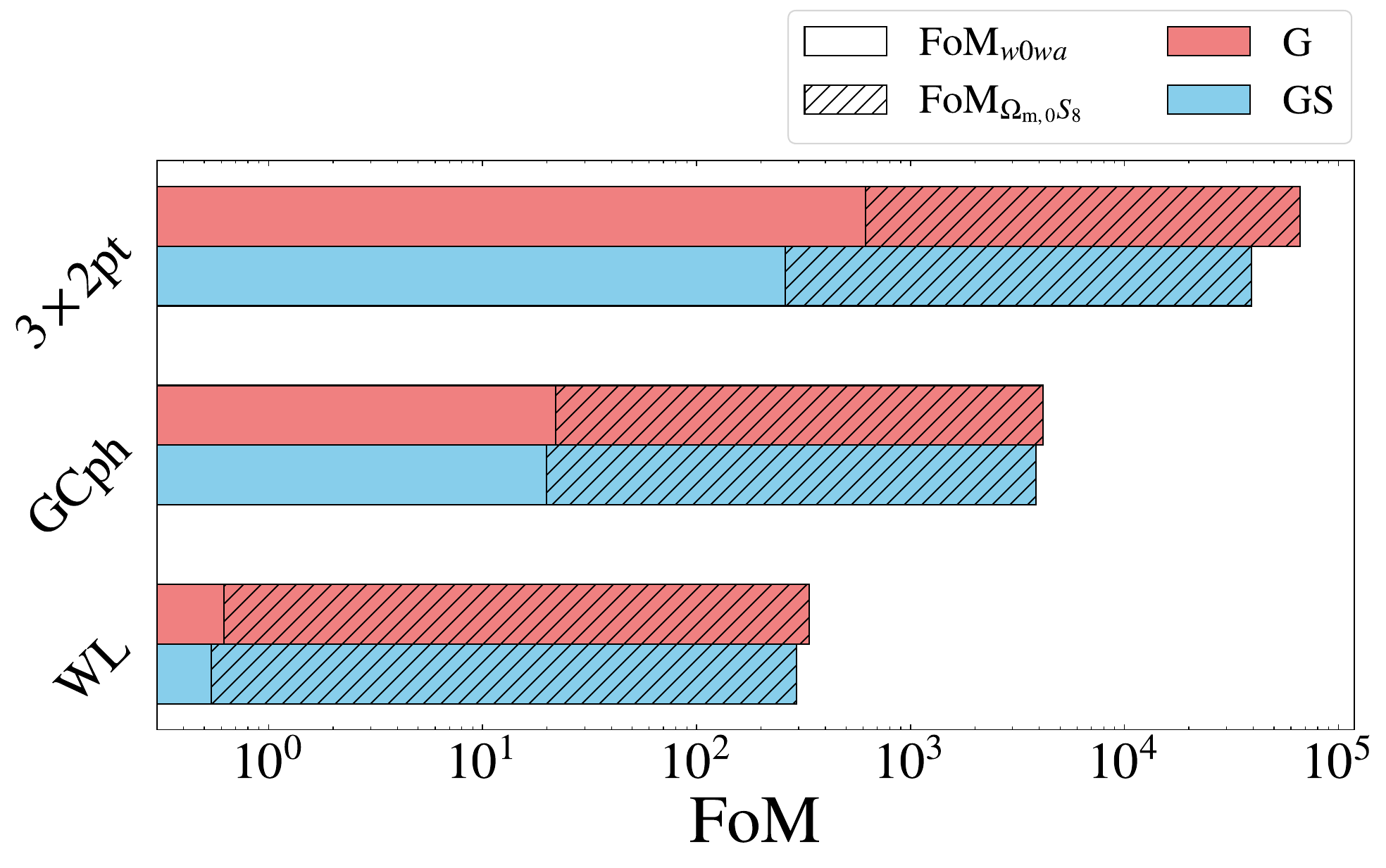}
    \caption{\dav{Same as Fig.~\ref{fig: barplot}, but for the dark energy and $\Omega_{\rm m, 0} - S_8$ FoM.}}
    \label{fig: barplot_fom}
\end{figure}
 \begin{figure}
    \includegraphics[width=0.48\textwidth]{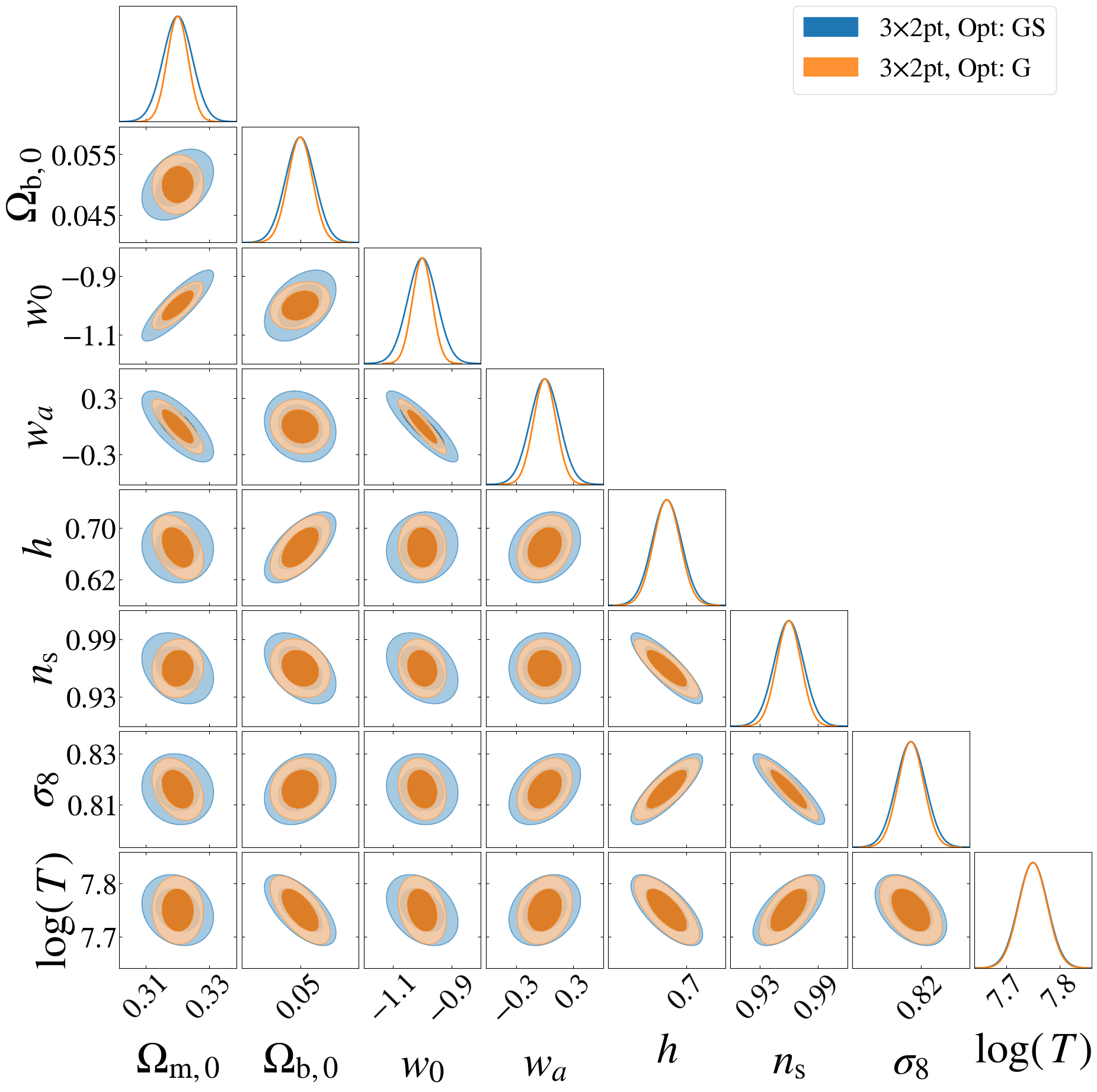}
    \caption{Contour plot for the G and GS constraints, considering the full 3$\times$2pt analysis in the optimistic case, in the reference scenario. \dav{The two shaded regions of the ellipses represent the 1 and 2$\sigma$ contours}. For clarity, the nuisance parameters are shown separately in Fig.~\ref{fig: nuisance} \dav{and the \logT\, parameter name has been shortened to $\log (T)$.}}
    \label{fig: triangle_plot}
 \end{figure}

\begin{table*}
\centering
\caption{\dav{Ratio between the GS and G marginalised uncertainties for all cosmological parameters and the FoM in the reference scenario, for the optimistic and pessimistic settings.}}
\begin{tabular}{l | c c c c c c c c | c c}
\hline
\multicolumn{1}{l |}{$\mathcal{R}(x)$} & $\Omega_{{\rm m},0}$ & $\Omega_{{\rm b},0}$ & $w_0$ & $w_a$ & $h$ & $n_{\rm s}$ & $\sigma_8$ & $\logten(T_{\rm AGN}/{\rm K})$ & ${\rm FoM}_{w0wa}$ & ${\rm FoM}_{\Omega_{{\rm m},0}S_8}$ \\
\hline
\hline
WL, Pes. & 
1.009 &   1.014 &   1.008 &   1.006 &   1.028 &   1.040 &   1.026 &   1.003 & 0.982 & 0.921 \\
WL, Opt. & 1.115 &   1.057 &   1.113 &   1.051 &   1.009 &   1.001 &   1.129 &   1.006 & 0.872 & 0.871 \\
\hline
\hline 
GCph, Pes. & 
 1.001 &   1.000 &   1.001 &   1.001 &   1.000 &   1.000 &   1.001 &   1.000 & 1.000 & 0.999\\
GCph, Pes., GCph diag. & 
1.000 &   1.000 &   1.000 &   1.000 &   1.000 &   1.000 &   1.000 &   1.000 &   1.000 & 1.000\\
GCph, Opt. & 
 1.030 &   1.006 &   1.091 &   1.091 &   1.033 &   1.003 &   1.060 &   1.004  & 0.910 & 0.924\\
GCph, Opt., GCph diag. & 
1.004 &   1.003 &   1.001 &   1.000 &   1.002 &   1.001 &   1.002 &   1.001 &   0.990 & 0.995\\
\hline   
\hline 
3$\times$2pt, Pes. & 
1.720 &   1.075 &   1.622 &   1.282 &   1.013 &   1.011 &   1.096 &   1.011  & 0.502 &  0.568\\
3$\times$2pt, Pes., GCph diag. & 1.673 &   1.092 &   1.463 &   1.202 &   1.016 &   1.011 &   1.113 &   1.009 &   0.541 &  0.584\\
3$\times$2pt, Opt. & 
1.401 &   1.197 &   1.489 &   1.301 &   1.102 &   1.206 &   1.143 &   1.034  & 0.422 &  0.593 \\
3$\times$2pt, Opt. , GCph diag.& 
1.401 &   1.205 &   1.422 &   1.255 &   1.106 &   1.215 &   1.146 &   1.030 &   0.437 & 0.589\\
\hline
\end{tabular}
\label{tab:ratio_ref_marginal}
\end{table*}

\dav{Having explored the impact of SSC on the conditional uncertainties, we move on to analyse a more realistic scenario, namely letting the parameters in the analysis free to vary, as opposed to fixing them to their fiducial values.\\
To compute the 1$\sigma$ uncertainties in this case, we marginalise over all cosmological and nuisance parameters. We add Gaussian priors of standard deviation $\sigma^{\rm p}=5\times 10^{-4}$ on the multiplicative shear bias parameters, and of $\sigma^{\rm p}=\sigma_z (1+z)$ with $\sigma_z = 0.002$ (see \citealt{Mellier2024} and references therein) on the $\diff z_i$ parameters. We also include a Gaussian prior of $\sigma^{\rm p} = 0.06$ on \logT\,, which roughly matches the prior range recommended in \cite{Mead2020}.
To add these priors in the FM analysis, it is sufficient to add $(\sigma_\alpha^{\rm p})^{-2}$ to the appropriate diagonal elements of the G and GS FMs ($\sigma_\alpha^{\rm p}$ being the value of the prior on parameter $\alpha$).}

\begin{figure*}
\centering
    \includegraphics[width=1.\textwidth]{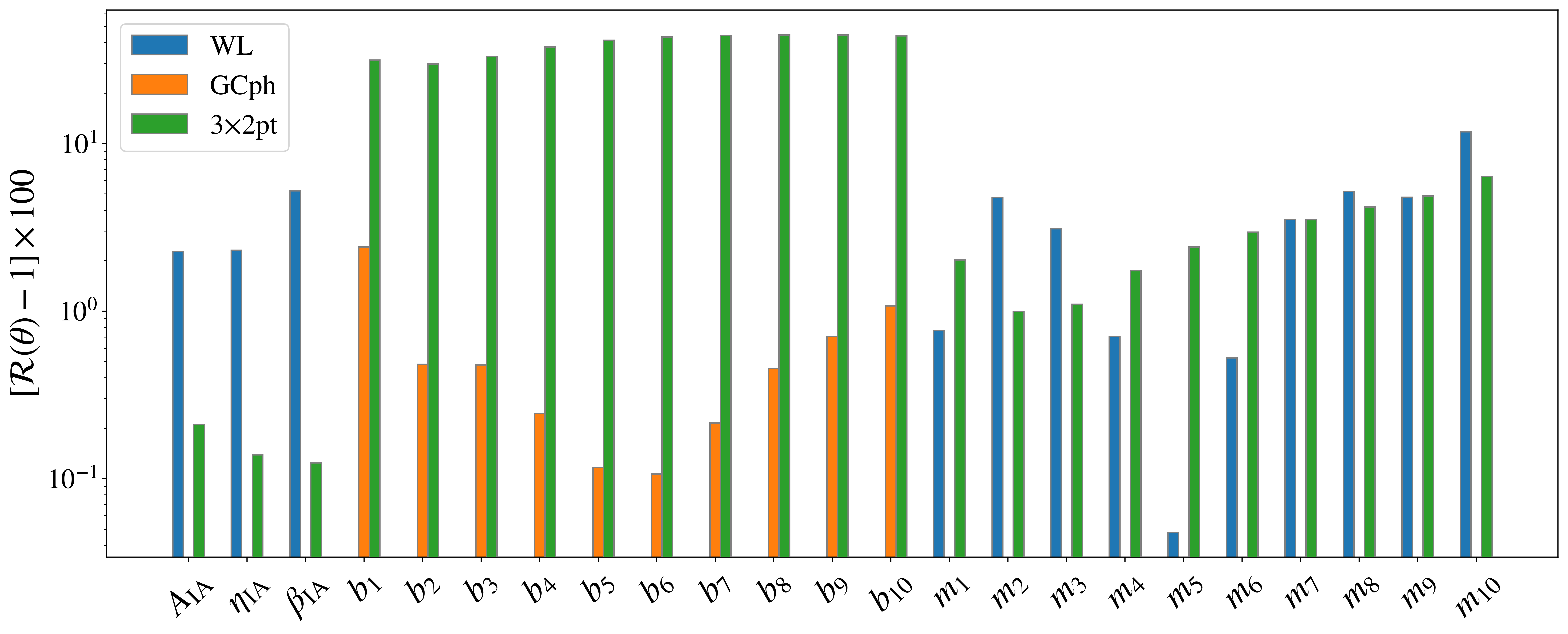}
    \caption{Percent increase of the marginalised $1\sigma$ uncertainty of the nuisance parameters, for all probe choices, in the optimistic case and for the reference scenario.}
    \label{fig: nuisance}
\end{figure*}

\dav{The results of this analysis, which we take as our reference scenario, are shown in Table~\ref{tab:ratio_ref_marginal} and in Figs.~\ref{fig: barplot} - \ref{fig: nuisance}. 
%In this case, the SSC impact on the WL cosmological constraints is dramatically reduced, amounting to an uncertainty increase of about 10\% at most in the optimistic case. The FoM also shows a similar reduction, related to the inclusion of non-linear modes that are more sensitive to the SSC and hence more pronounced for the optimistic case. This is one of the main results of our paper: including and marginalising over nuisance parameters ...\\
For the WL case, marginalisation over all cosmological and nuisance parameters leads to the result that the SSC has now a very minor impact -- of a maximum of about 13\% in the optimistic case -- on the constraints and the FoM: again, this is in line with what found in \cite{Barreira2018cosmic_shear}. We also compute the $\Omega_{{\rm m},0} - S_8$ FoM, after projecting the FM to the new parameter space with $S_8$ replacing $\sigma_8$ (see e.g. \citealt{Coe2009_FM}).
This drops from 336 to 292 in the optimistic case, corresponding to a ratio of 0.87 (0.92 in the pessimistic case), closely mirroring the results found for ${\rm FoM}_{w0wa}$.\\ 
These values are actually easily explained: marginalising over cosmological and nuisance parameters, particularly if these are degenerate with the amplitude of the signal, dilutes the SSC effect in a larger error budget; because of this, it is the relative rather than the absolute impact of SSC that decreases. Indeed, marginalising over additional parameters is formally equivalent to having additional covariance.
The parameters that mostly change the SSC impact when marginalised over are the cosmological ones, but the multiplicative nuisance parameters $m_i$ and $b_i$ also play a role: adding these to the set of free parameters introduces a degeneracy between these and the overall amplitude of $C_{ij}^{\rm AB}(\ell)$. Such a degeneracy is a mathematical one present on the whole $\ell$ range. As a consequence, the constraints on all the parameters and the FoM are degraded in a way that is independent of the presence of SSC. This is shown in Figs.~\ref{fig: barplot} and \ref{fig: barplot_fom}, which respectively exhibit the relative uncertainty $\bar{\sigma}$ and the FoMs in the G and GS cases for each parameter, if we marginalise or not over nuisance parameters. Letting these free to vary, i.e. marginalising over them, tends to increase the uncertainty on cosmological parameters way more than including SSC, and this is even more true when these nuisance parameters are simply multiplicative such as $b_i$ and $m_i$.\\}
This of course does not mean that varying more parameters improves the constraints. Indeed, the uncertainties on all parameters increase (hence the FoM decreases) with respect to the case of conditional uncertainties introduced above. \\
%\sout{The degradation is, however, the same with and without SSC so the ${\cal{R}}(\theta)$ values stay close to unity}.
\indent
\dav{The same reasoning applies to the GCph probe, for which the SSC impact drops now to a subpercent level in the pessimistic case and to about 10\% at most in the optimistic case (on $w_0$ and the two FoMs). As mentioned above, for GCph, one of the reasons behind the observed decrease in the (already low) SSC relative impact is the marginalisation over the galaxy bias parameters, which are perfectly degenerate with the amplitude of the signal and over which we impose no prior.}\\
\indent
\dav{On the other hand, the results for the 3$\times$2pt case show that the SSC still matters. The additional information carried by the GCph and XC data allows the partial breaking of parameter degeneracies, including those with probe-specific systematics such as $m_i$ and $b_i$, hence making the scale-dependent increase of the uncertainties due to the inclusion of SSC important again. For this reason, in this case, the 3$\times$2pt does not follow the behaviour of the single probes as seen earlier. In particular, the dark energy FoM, whose increase with respect to present surveys is one of the main objectives of the \Euclid mission, is highly degraded -- by a factor of about 2 in the optimistic case. This is mainly due to the large impact on the dark energy equation of state parameters, showing the importance of accounting for SSC in upcoming LSS analyses. The same conclusion holds for ${\rm FoM}_{\Omega_{\rm m, 0}S_8}$, with a slightly larger $\cal{R}$ value driven by the lower impact on $\sigma_8$ w.r.t. $w_0$ and $w_a$. In Fig.~\ref{fig: triangle_plot} we show the comparison of the 2D contours for all cosmological parameters between G and GS in the case of the 3$\times$2pt analysis, in the optimistic case; the most impacted parameters in this case are $\Omega_{{\rm m},0}, w_0$ and $w_a$. In addition, this shows that SSC does not seem to strongly affect the correlations between cosmological parameters.}\\
\indent
\dav{Table~\ref{tab:ratio_ref_marginal} also shows the ratios in the `GCph diag.'} case, in which we neglect the GCph cross-reshift elements (i.e. we take the diagonal of the $C_{ij}^{\rm GG}(\ell)$ matrix for each $\ell$ value). This choice is sometimes made because of the large sensitivity of such measurements on the photometric redshift calibration, which is less of an issue for WL due to its broad kernel. The values show the robustness of our conclusions even in this case, with the $\cal{R}$ values being very close to the standard case both for GCph and for the 3$\times
$2pt.\\
\indent
\dav{We note that these conclusions depend on the scale cuts imposed in the data vector, as can be seen from the lower FoM decrease (and of the SSC impact as a whole, with some exceptions) in the pessimistic case, in line with what found in \citet{lacasa_20} for GCph. 
This is a direct consequence of the larger amount of non-linear modes included in the data vectors, which as mentioned previously are more subject to mode coupling and hence contribute more to SSC. For WL, it should be noted that the diagonal elements of the total covariance matrix are always dominated by the Gaussian contribution, because of the presence of the scale-independent shape noise (see Eq.~\ref{eq: covgauss} for $A = B = {\rm L}$), which largely dominates over the SSC on small scales. This is consistent with the results of \citet{upham2021} showing that the diagonal elements of the WL total covariance matrix are more and more dominated by the Gaussian term as we move to higher $\ell$. This is also the case for GCph, although the predominance of the Gaussian term along the diagonal is less pronounced because of the smaller contribution of shot noise.\\
As mentioned above, more sophisticated choices of scale cuts, e.g. through the use of the BNT transform, will allow decreasing the high-$k$ contribution to a given $\ell$ mode, hence mitigating the SSC impact more than the \enquote{hard} angular scale cut considered here (and in \citetalias{ISTF2020}). We leave the investigation of this point to a forthcoming publication.}

To conclude this section, it is also worth looking at the impact of SSC on the astrophysical nuisance parameters. Indeed, although an issue to be marginalised over when looking at cosmological ones, the IA and the galaxy bias parameters are of astrophysical interest. \dav{We show the impact of SSC on the constraints on these quantities in Fig.~\ref{fig: nuisance}, as well as on the galaxy bias and the multiplicative shear bias parameters}. \\
\indent
For IA-related nuisance parameters, the uncertainty increase due to SSC is lower than \dav{5\% when considering WL-only, and below 1\% for the full case. The multiplicative shear bias parameters are affected by a similar amount (up to around 10\% for the last bin) and in a similar way for WL and the 3$\times$2pt. As for the galaxy bias parameters, the impact is modest for GCph but quite significant (between 30 and 40\%) for the 3$\times$2pt. These results are analogous to what was found before for the cosmological parameters: in the marginalised 3$\times$2pt case the impact of SSC is more apparent, since many degeneracies are broken thanks to the probes combination and cross-correlation. \\
In this case, contrary to the IA parameters, the uncertainty on $b_i$ and $m_i$ in each of the ten redshift bins is significantly affected by SSC. This is because both of these nuisance parameters simply act as a multiplicative factor on the power spectrum and are thus highly degenerated with the effect of SSC. Again, this is related to the fact that the first-order effect of SSC is to modulate the overall clustering amplitude because of a shift in the background density $\delta_{\rm b}$. 
For the $m_i$ parameters, we remind that a tight (while realistic) prior is imposed, keeping the SSC uncertainty increase, and hence the $\cal{R}$ ratio, quite low even for the full 3$\times$2pt.}
We note that going beyond the linear approximation for the modelling of the galaxy bias will add more nuisance parameters, thus degrading the overall constraints on cosmological parameters and further reducing the relative impact of SSC.

%\textcolor{green}{Finally, comparing how uncertainties on $b_i$ and $m_i$ react to the addition of SSC, we can see that surprisingly the $b_i$ are more affected in the 3$\times$2pt case than in the GCph case, while it is the contrary for $m_i$, the uncertainty increase is larger for WL than for 3$\times$2pt. This difference in the behaviour of the uncertainty increase might come from the numerous degeneracies existing between these nuisance parameters and the most constrained cosmological parameters in each case. Though it is not easy to exactly understand this behaviour, we note that in all cases the $\mathcal{R}(\theta)$ for these parameters are of the same order of magnitude and are never completely negligible.}
%
\begin{table*}
\centering
\caption{Same as Table~\ref{tab:ratio_ref_marginal} but removing the flatness prior.}
\begin{tabular}{l | c c c c c c c c c | c c }
\hline
\multicolumn{1}{l |}{$\mathcal{R}(x)$} & $\Omega_{{\rm m},0}$ & $\Omega_{{\rm DE, 0}}$ & $\Omega_{{\rm b},0}$ & $w_0$ & $w_a$ & $h$ & $n_{\rm s}$ & $\sigma_8$ & $\logten(T_{\rm AGN}/{\rm K})$ & ${\rm FoM}_{w0wa}$ & ${\rm FoM}_{\Omega_{\rm m, 0}S_8}$\\
\hline
\hline
\multicolumn{1}{l |}{WL, Pes.} & 
1.022 & 1.002 & 1.012 & 1.019 & 1.009 & 1.001 & 1.012 & 1.003 & 1.005 & 0.979 & 0.966 \\
\multicolumn{1}{l |}{WL, Opt.} & 
1.059 & 1.001 & 1.016 & 1.060 & 1.019 & 1.010 & 1.004 & 1.058& 1.001 & 0.937 & 0.945 \\
\hline
\hline
\multicolumn{1}{l |}{GCph, Pes.} & 
1.004 & 1.002 & 1.003 & 1.001 & 1.003 & 1.001 & 1.002 & 1.001 & 1.000 & 0.995 & 0.996\\
\multicolumn{1}{l |}{GCph, Opt.} & 
1.035 & 1.026 & 1.017 & 1.062 & 1.062 & 1.021 & 1.012 & 1.037 & 1.008  & 0.890 & 0.908 \\
\hline
\hline
\multicolumn{1}{l |}{3$\times$2pt, Pes.} & 
1.871 & 1.227 & 1.129 & 1.399 & 1.343 & 1.031 & 1.019 & 1.310 & 1.010 &  0.482 & 0.534\\
\multicolumn{1}{l |}{3$\times$2pt, Opt.} & 
1.491 & 1.126 & 1.117 & 1.462 & 1.207 & 1.022 & 1.081 & 1.060 & 1.083 & 0.548 & 0.655\\
\hline
\end{tabular}
\label{tab:ratio_nonflat}
\end{table*}
\subsection{Non-flat cosmologies}
In the previous section, we investigate the SSC on the cosmological parameters under the assumption of a flat model. Actually, the requirement on the FoM assessed in the \Euclid Red Book \citep{laureijs2011euclid} refers to the case with the curvature as an additional free parameter to be constrained, that is, the non-flat $w_0w_a$CDM model. This is why in \citetalias{ISTF2020} are also reported the marginalised uncertainties for the parameter $\Omega_{{\rm DE,0}}$, with a fiducial value $\Omega_{{\rm DE,0}}^{\rm fid} = 1 - \Omega_{{\rm m,0}}$ to be consistent with a flat universe. It is then worth wondering what the impact of SSC is in this case too. This is summarised in Table~\ref{tab:ratio_nonflat}, where we now also include the impact on $\Omega_{{\rm DE,0}}$. \\
\indent
A comparison with the results in Table~\ref{tab:ratio_ref_marginal} is quite hard if we look at the single parameters. Indeed, opening up the parameter space by removing the flatness assumption introduces additional degeneracy among the parameters controlling the background expansion, which are thus less constrained whether SSC is included or not. \dav{We can nevertheless note again that, for the marginalised uncertainties, 3$\times$2pt is still the most impacted probe; the difference between pessimistic and optimistic scenarios is now less evident with ${\cal{R}}(\theta)$ increasing or decreasing depending on the parameter and probe considered.}\\
Once more, the most affected parameters for WL are $(\Omega_{{\rm m,0}}, \sigma_8)$, the uncertainties on which are now further degraded by the fact that they correlate with the parameter $\Omega_{{\rm DE,0}}$ which is also affected. Although $(w_0, w_a)$ are also degraded by the SSC, a sort of compensation is at work, so that the overall decrease in the FoM is similar to the case with the flatness prior. The motivations that make GCph much less affected still hold when dropping the flatness prior, explaining the corresponding $\mathcal{R}(\theta)$ values.

\dav{We also note a slight increase of ${\cal{R}}({\rm FoM})$ in the 3$\times$2pt optimistic case, meaning a smaller degradation of the FoM due to SSC. 
The dark energy FoM indeed degrades by $52\%$ $(45\%)$ in the non-flat case vs. $50\%$ $(58\%)$ for the flat case in the pessimistic (optimistic) scenario,
while ${\rm FoM}_{\Omega_{\rm m, 0}S_8}$ degrades by $47\%$ $(35\%)$ in the non-flat case vs. $43\%$ $(41\%)$ for the flat case in the pessimistic (optimistic) scenario}. 
This can be qualitatively explained by noting that the decrease of both FoM(G) and FoM(GS) is related to a geometrical degeneracy which is the same on all scales, whether or not they are affected by the increase in uncertainty due to the SSC inclusion.\\

Overall, these results suggest a dependence of the SSC significance on both the number and type of parameters to be constrained. Qualitatively, we can argue that SSC is more or less important depending on whether the additional parameters (with respect to the reference case of a flat model) introduce degeneracies which are or not scale-dependent and how strong is the degeneracy between these parameters and the amplitude of the power spectrum. To give an example, in future works lens magnification effects should be included in the analysis as these were shown to have a significant impact on cosmological constraints \citep{Unruh_19}. From our results, we can anticipate that the inclusion of magnification-related nuisance parameters will further dilute the impact of SSC.

\subsection{Dependence on redshift binning}
\label{sect: red_bin}
\begin{figure*}[!ht]
\centering
\includegraphics[width=\textwidth]{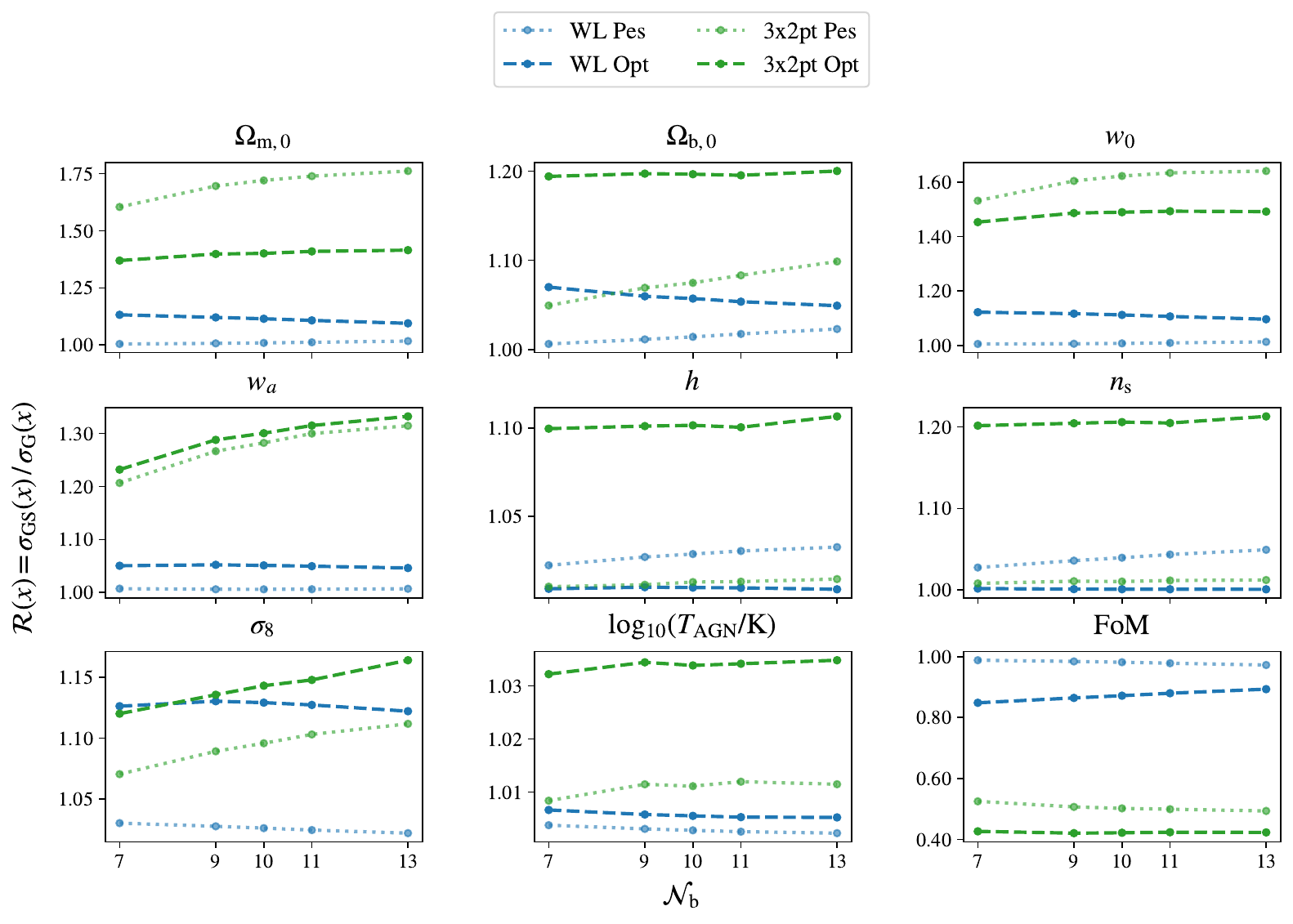}
\caption{Ratio between WL and 3$\times$2pt marginalised uncertainties computed by including or neglecting the SSC contribution, as a function of the number of (equally populated) redshift bins, for the pessimistic and optimistic cases.}
\label{fig: ratiovsnbin}
\end{figure*}
The results summarised in Tables\,\ref{tab:ratio_ref_conditional}--\ref{tab:ratio_nonflat} were obtained for a fixed choice of number and type of redshift bins. We investigate here how the results depend on these settings given that we expect both the G and GS constraints to change as we vary the number and type of bins. We consider again the reference scenario introduced in Sect.~\ref{sec:reference_scenario}, that is, the case of \dav{flat models, marginalising over the full set of nuisance parameters while imposing Gaussian priors over some of them. In this section we only consider the WL and 3$\times$2pt cases, since as we have seen SSC has a modest impact on GCph.}

Let us first consider changing the number of redshift bins ${\cal N}_{\rm b}$. We show the scaling of ${\cal{R}}(\theta)$ as a function of ${\cal N}_{\rm b}$ for the WL and 3$\times$2pt probes, respectively, in Fig.~\ref{fig: ratiovsnbin} -- for both the pessimistic and optimistic assumptions. \dav{The most remarkable result is the weak dependence of ${\cal{R}}(x)$ w.r.t. ${\cal N}_{\rm b}$, as can be inferred from the small $y$ range spanned by the curves -- see e.g. the bottom right panel, for the FoM}. More specifically, the scaling of ${\cal{R}}(\theta)$ with ${\cal N}_{\rm b}$ depends on the parameter and the probe one is looking at. It is quite hard to explain the observed trends because of the interplay of different contrasting effects. For instance, a larger number of bins implies a smaller number density in each bin, and hence a larger shot noise. As a consequence, the SSC contribution to the total covariance for the diagonal elements will likely be more and more dominated by the Gaussian component because of the larger shot and shape noise terms. However, this effect also depends on the scale so that, should the SSC be the dominant component on the scales to which a parameter is most sensitive, the impact should still be important. On the other hand, a larger number of bins also comes with a larger number of nuisance parameters which, as shown above, leads to a reduction of the SSC impact. Quantifying which actor plays the major role is hard which explains the variety of trends in the different panels.

\begin{figure*}[!ht]
\centering
\includegraphics[width=\textwidth]{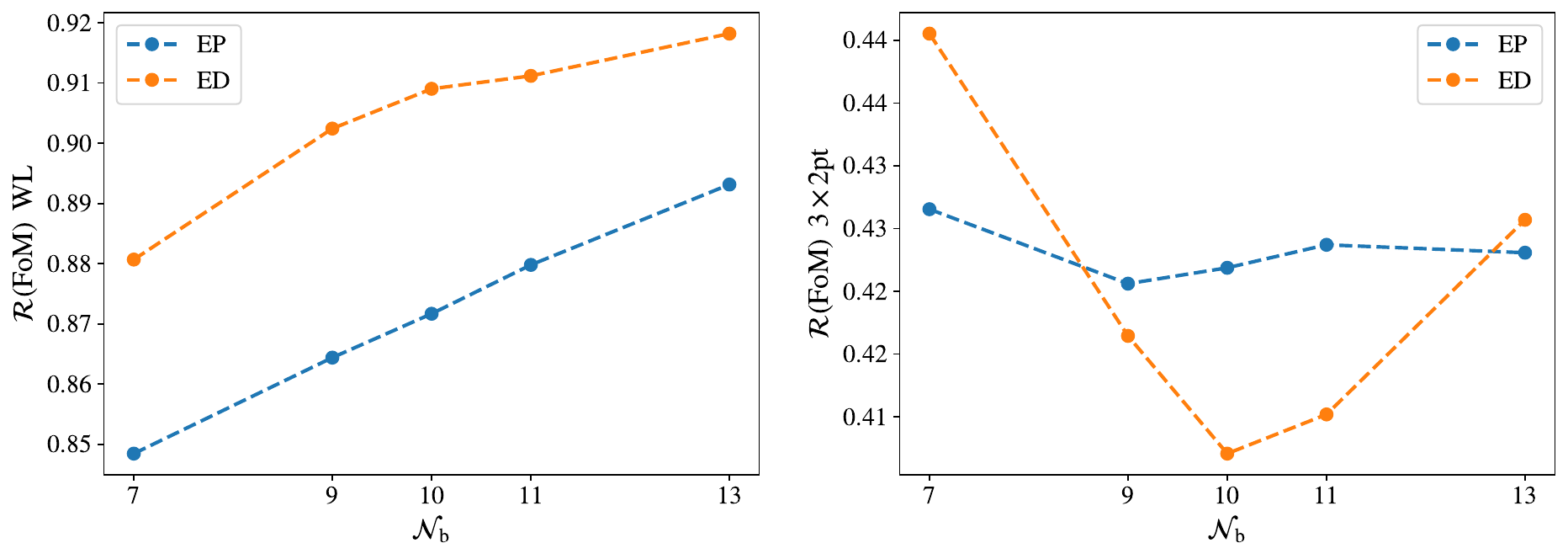}
\caption{FoM ratio vs the number of EP and ED redshift bins for WL (left) and 3$\times$2pt (right) in the optimistic scenario. \dav{The blue curves in this plot match the ones in the bottom right panel of Fig~\ref{fig: ratiovsnbin} in the optimistic case, for the corresponding probes.}}
\label{fig: ratioeped}
\end{figure*}
\begin{figure*}[!ht]
\centering
\includegraphics[width=\textwidth]{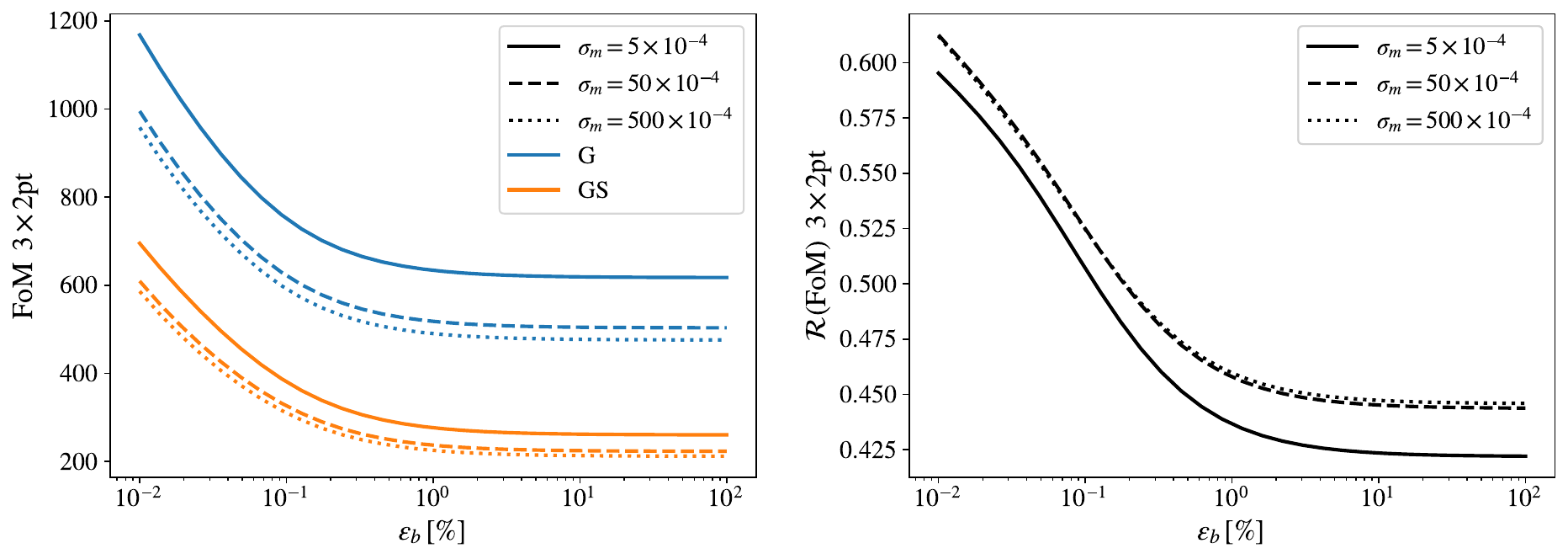}
\caption{\davthree{3$\times$2pt FoM in the optimistic scenario with and without SSC as a function of the percentage prior $\varepsilon_b$ on the galaxy bias parameters for $\sigma_m = (5, 50, 500) \times 10^{-4}$ (solid, dashed, dotted lines). 
\textit{Left}: FoM values. {\it Right.} ${\cal R}({\rm FoM})$, defined as the ratio between the blue and red curves.}}
\label{fig: ratiofomvsprior}
\end{figure*}
As a further modification to the reference settings, we can change how the redshift bins are defined. We have up to now considered equipopulated (EP) bins so that the central bins cover a smaller range in $z$, because of the larger source number density. As an alternative, we divide the full redshift range into ${\cal N}_{\rm b}$ bins with equal \dav{redshift support (`{equidistant}', ED)}, and recompute the FM forecasts with and without SSC. We show the FoM ratio as a function of the number of bins for EP and ED bins considering WL (left) and 3$\times$2pt (right) probes in the optimistic scenario in Fig.~\ref{fig: ratioeped}. We note that finding the exact number and type of redshift bins used to maximise the constraining power of \Euclid is outside the scope of this paper; this effort is indeed brought forward in the context of the SPV exercise.

In order to qualitatively explain these results, let us first consider the WL case. Given that the bins are no longer equipopulated, the number density of galaxies will typically be larger in the lower redshift bins than in the higher ones. As a consequence, the larger the number of bins, the higher the shape noise in the higher redshift bins so that the SSC will be subdominant in a larger number of bins, which explains why its impact decreases (i.e. ${\cal{R}}({\rm FoM})$ increases) with ${\cal N}_{\rm b}$. Nevertheless, the impact of SSC will be larger than in the EP case since SSC will dominate in the low redshift bins which are the ones with the largest \dav{signal-to-noise ratio}. \dav{This effect is small, and the difference in  ${\cal{R}}({\rm FoM})$ between ED and EP is no larger than 3--5\%.}

When adding GCph and XC into the game, the impact of SSC is determined by a combination of contrasting effects. On the one hand, we can repeat the same qualitative argument made for WL also for GCph and XC\footnote{\dav{We note that, although the $C(\ell)$ for XC are not affected by noise, (cfr. Eq.~\ref{eq: noiseps}), their covariance is (cfr. Eq.~\ref{eq: covgauss}).}} thus pointing at ${\cal{R}}({\rm FoM})$ increasing with ${\cal N}_{\rm b}$. The larger the number of bins, the narrower they are, and the smaller the cross-correlation between them hence the smaller the Gaussian covariance. This in turn increases the number of elements in the data vector whose uncertainty is dominated by the SSC. Should this effect dominate, we would observe a decrease of ${\cal{R}}({\rm FoM})$ with ${\cal N}_{\rm b}$ with the opposite trend if it is the variation of the shape and shot noise to matter the most. This qualitative argument allows us then to roughly explain the non-monotonic behaviour of ${\cal{R}}({\rm FoM})$ we see in the right panel of Fig.~\ref{fig: ratioeped}. \\
It is worth remarking, however, that the overall change of ${\cal{R}}({\rm FoM})$ for subsequent ED (or EP) bins over the range in ${\cal N}_{\rm b}$ is smaller than \dav{$\sim 5\%$} which is also the \dav{maximum} value of the difference between ${\cal{R}}({\rm FoM})$ values for EP and ED bins once ${\cal N}_{\rm b}$ is fixed. 

The analysis in this section motivates us to argue that the constraints and FoM degradation due to SSC \dav{have a weak dependence on the redshift binning scheme}.
\begin{figure}[ht]
\centering
\includegraphics[width=\hsize]{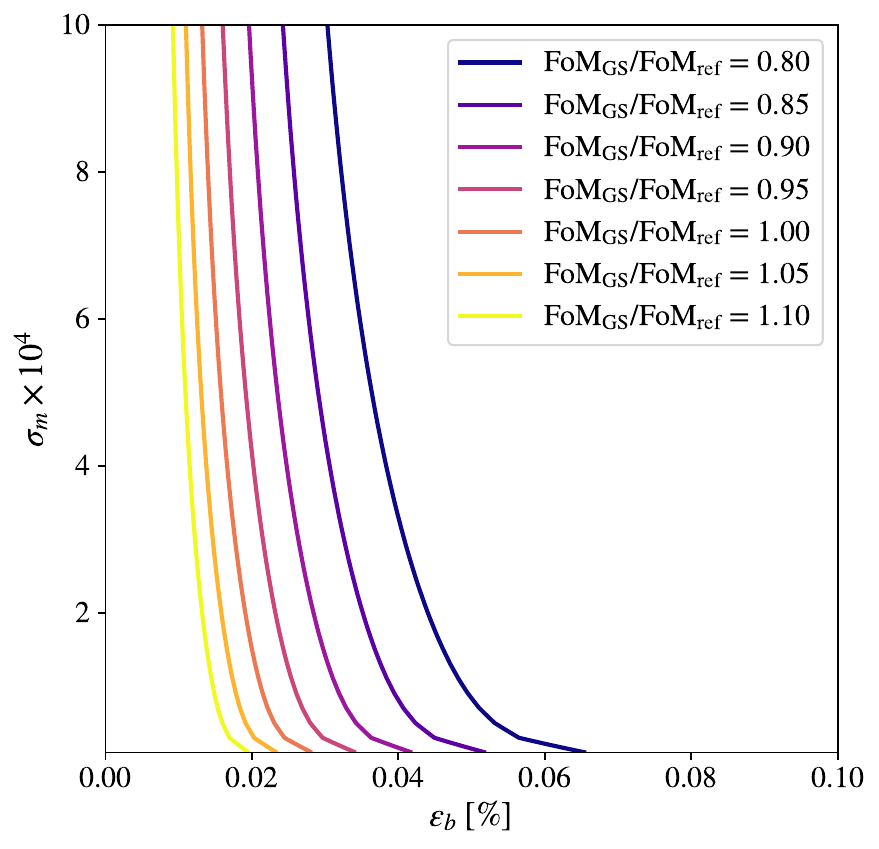}
\caption{${\rm FoM_{GS}}$ contours in the $(\varepsilon_b, \sigma_m)$ plane for ${\rm FoM_{GS}/FoM_{\rm ref}}$ going from 0.8 to 1.1 in steps of 0.05 (from right to left).} 
\label{fig: gsrefvsprior}
\end{figure}
\subsection{Requirements on prior information}

The results in the previous paragraph show that the SSC may dramatically impact the constraints on the cosmological parameters. As a consequence, the 3$\times$2pt FoM is reduced by up to \dav{$\sim 50\%$} with respect to the case when only the Gaussian term is included in the total covariance. This decrease in the FoM should actually not be interpreted as a loss of information due to the addition of the SSC. On the contrary, one can qualitatively say that removing SSC from the error budget is the same as adding information that is not actually there. 
%As such, one should not attempt to recover the Gaussian FoM simply because that value overestimated the constraints the data can put on the cosmological parameters. 
It is nevertheless interesting to ask which additional information must be added to recover the Gaussian FoM, which is usually taken as a reference for gauging the potential of a survey. 
%This information can only come from priors on the nuisance (or cosmological) parameters since there is no way to retrieve them from data affected by the SSC. 
This information can come from priors on the nuisance (or cosmological) parameters. 
In the following section, we investigate the former option by adding Gaussian priors on the galaxy and multiplicative shear bias parameters. \\
\indent
To this end, we consider the realistic case of a \dav{flat} model plus the galaxy bias and multiplicative shear bias as nuisance parameters. As a simplifying assumption, we assume that all the ${\cal N}_{\rm b}$ bias values $b_i$ are known with the same percentage uncertainty $\varepsilon_b = \sigma_{b, \, i}/b_{{\rm fid}, \, i}$, while we put \dav{the same} prior $\sigma_m$ on all the $m_i$ parameters (having set the fiducial value $m_{\rm fid}$ to 0). We then compute the FoM with and without SSC for the 3$\times$2pt probe in the optimistic scenario and investigate how the ratio ${\cal{R}}({\rm FoM})$ scales with $(\varepsilon_b, \sigma_m)$ obtaining the results shown in Fig.~\ref{fig: ratiofomvsprior}.

\dav{Both in the G and GS cases, the FoM shows little to no sensitivity to priors above $\sim 1\%$ and no sign of saturation even for extremely narrow priors (as little as 0.01\%) because of the presence of such a large number of nuisance parameters. For $\sigma_m$ we observe a similar behaviour, with the difference between the curves corresponding to the two larger priors values being visibly smaller than the difference between the curves corresponding to the smaller ones. This is less true for the GS FoM (orange curves), as the presence of SSC in the error budget decreases the relative overall improvement coming from a smaller uncertainty over the $m_i$ parameters.\\}
\indent
A prior on the nuisance parameters increases both the Gaussian and Gaussian\,+\,SSC FoM so that one could expect their ratio to be independent of the prior itself. This is not exactly the case since the correlation between different multipoles introduced by SSC alters the way the prior changes the FM elements. As a result, we find a non-flat scaling of ${\cal{R}}({\rm FoM})$ as can be seen from the right panel of Fig.~\ref{fig: ratiofomvsprior}. 
%When a strong prior is set on the galaxy bias (i.e. $\varepsilon_b \ll 1$), there is not much gain in improving the knowledge of the multiplicative shear bias so that the solid, dashed, and dotted lines (corresponding to three $\sigma_m$ values) are quite close to each other. This is no longer the case for larger $\varepsilon_b$ values (i.e. weak or no prior on the bias): lowering $\sigma_m$ has now a larger impact on ${\cal{R}}({\rm FoM})$.
%although a saturation is quickly reached causing the dashed and dotted lines to be closer than the solid and dashed ones. 
\dav{The behaviour of ${\cal{R}}({\rm FoM})$ with $\varepsilon_b$ tells us that ${\rm FoM}_{\rm GS}$ increases with decreasing $\varepsilon_b$ slower than ${\rm FoM}_{\rm G}$ when the galaxy bias is known with an uncertainty smaller than the percent level. Another way to interpret it is that the information gained in the FoM saturates faster when SSC is included: better constraints on $\varepsilon_b$ do not bring more information as the SSC now dominates the error budget. However, it is worth stressing that, even for a strong prior on the multiplicative shear bias, the FoM ratio can actually be improved significantly only under the (likely unrealistic) assumption of a subpercent prior on the galaxy bias.}

The need for such strong priors comes from the attempt to retrieve the same FoM as a Gaussian case. Alternatively, one can also wonder which additional information must be added through priors to retrieve the idealised FoM value obtained in forecasts that neglect the SSC. In other words, we look for the requirements that must be put on the priors $(\varepsilon_b, \sigma_m)$ in order to make ${\rm FoM}_{\rm GS}/{\rm FoM}_{\rm ref} = 1$, where \dav{${\rm FoM}_{\rm ref} = 617$} is the FoM computed for a flat reference case without SSC and with no priors on galaxy bias, but a fiducial prior $\sigma_m = 5 \times 10^{-4}$ on the shear bias. The answer to this question is shown in Fig.~\ref{fig: gsrefvsprior} for the optimistic scenario and 10 equipopulated redshift bins. Some numbers help to better understand how priors can indeed supply the additional information to retrieve the FoM one would obtain in an ideal case where SSC is absent. Solving 
%Using the same prior $\sigma_m$ for both the reference and the SSC FoM and solving
%
\begin{displaymath}
{\rm FoM}_{\rm GS}(\varepsilon_b, \sigma_m) = f \, {\rm FoM}_{\rm ref}
\end{displaymath}
with respect to $\varepsilon_b$, we get
\begin{displaymath}
    \varepsilon_b = \left \{
    \begin{array}{ll}
    \displaystyle{(0.05, 0.03, 0.02) \, \%} & \displaystyle{{\rm for} \; \sigma_m = 0.5 \times 10^{-4}} \\
    & \\
    \displaystyle{(0.04, 0.02, 0.02) \, \%} & \displaystyle{{\rm for} \; \sigma_m = 5 \times 10^{-4}} \\
    & \\
    \displaystyle{(0.03, 0.02, 0.01) \, \%} & \displaystyle{{\rm for} \; \sigma_m = 10 \times 10^{-4}} \; , \\
    \end{array}
    \right . 
\end{displaymath}
%
%
%    old, for sigmam = (5e-4, 5e-3, 5e-2)
%    \begin{displaymath}
%    \varepsilon_b = \left \{
%    \begin{array}{ll}
%    \displaystyle{(2.27, 1.18, 0.85)\%} & \displaystyle{{\rm for} \; \sigma_m = 5 \times 10^{-4}} \\
%    & \\
%    \displaystyle{(1.58, 1.00, 0.76)\%} & \displaystyle{{\rm for} \; \sigma_m = 5 \times 10^{-3}} \\
%    & \\
%    \displaystyle{(1.07, 0.78, 0.62)\%} & \displaystyle{{\rm for} \; \sigma_m = 5 \times 10^{-2}} \; , \\
%    \end{array}
%    \right . 
%    \end{displaymath}

%
where the three values refer to $f = (0.8, 0.9, 1.0)$. These numbers (and the contours in Fig.~\ref{fig: gsrefvsprior}) show that \dav{it is possible} to compensate for the degradation due to SSC \dav{only} by adding strong priors on the galaxy bias, which have a much larger impact on the (G and GS) FoM than strong priors on the multiplicative shear bias. However, it is worth noticing that it is actually easier to obtain priors on the multiplicative shear bias provided a sufficient number of realistic image simulations are produced and fed to the shear measurement code to test its performance. It is therefore worth wondering how much the FoM is restored by improving the prior on $m$ for a fixed one on the bias. We find
%
% \begin{comment}
%     old values, for $\sigma_m = (0.5, 5.0, 50) \times 10^{-4}$
    
%     \begin{displaymath}
%     \frac{{\rm FoM_{GS}}}{{\rm FoM_{ref}}} = \left \{
%     \begin{array}{ll}
%     \displaystyle{(2.87, 2.86, 2.71)} & \displaystyle{{\rm for} \; \varepsilon_b = 0.1\%} \\
%      & \\
%     \displaystyle{(0.95, 0.94, 0.90)} & \displaystyle{{\rm for} \; \varepsilon_b = 1\%} \\
%      & \\
%     \displaystyle{(0.76, 0.75, 0.72)} & \displaystyle{{\rm for} \; \varepsilon_b = 10\%} \; , \\
%     \end{array}
%     \right .
%     \end{displaymath}
% \end{comment}
%
\begin{displaymath}
    \frac{{\rm FoM_{GS}}}{{\rm FoM_{ref}}} = \left \{
    \begin{array}{ll}
    \displaystyle{(1.24, 1.13, 1.08)} & \displaystyle{{\rm for} \; \varepsilon_b = 0.01\%} \\
     & \\
    \displaystyle{(0.68, 0.62, 0.59 )} & \displaystyle{{\rm for} \; \varepsilon_b = 0.1\%} \\
     & \\
    \displaystyle{(0.5,  0.45, 0.43)} & \displaystyle{{\rm for} \; \varepsilon_b = 1\%} \; , \\
    \end{array}
    \right .
    \end{displaymath}
with the three values referring to $\sigma_m = (0.5, 5.0, 10) \times 10^{-4}$. As expected, improving the prior on the multiplicative bias with respect to the fiducial one (which, we remind, is included in ${\rm FoM_{\rm ref}}$) does not help a lot in recovering the constraining power. \dav{A very tight prior of around $0.1\%$ prior on the galaxy bias can recover a significant amount of the reference FoM (almost 70\%) thanks to the additional information compensating for the presence of SSC}. 

Investigating whether the priors proposed here can be achieved in practice (e.g. through theoretical bias models tailored to galaxy clustering data or N-body hydrodynamic simulations) is outside the aim of this work. We refer the interested reader to for example \citet{Alex2021}, \citet{Zen22}, and \citet{Ivanov2024} for some preliminary results.

\section{Conclusions}\label{sec: conclu}

Precision cosmology requires precision computation: previously neglected theoretical contributions must therefore now be taken into account. Motivated by this consideration, we computed and studied the impact of SSC on the \Euclid photometric survey, exploring how the different probes and their combination are affected by this additional, non-Gaussian term in the covariance matrix. The analysis of the impact of SSC on the spectroscopic survey, which has been shown to be small in \cite{wadekar_20} for the Baryon Oscillation Spectroscopic Survey (BOSS)  data, is left for future work.  We employed a FM analysis, producing forecasts of the $1\sigma$ marginalised uncertainties on the measurement of the cosmological parameters of the flat and non-flat $w_0w_a$CDM cosmological models. We validated two different forecast pipelines against the results of \citetalias{ISTF2020}, taking as reference survey the one specified therein, and then updated the galaxy bias and the source redshift distributions according to the most recent versions presented in \citet{Pocino2021}. The SSC was computed relying on the analytical approximations and numerical routines presented in \citetalias{Lacasa_2019}, interfacing the public code \texttt{PySSC} with two distinct forecast pipelines to validate the constraints. As a further step forward, we build upon the work of \citetalias{Lacasa_2019} by computing the scale and redshift dependence of the response functions of the different probes, starting from the results of \citet{Wagner2015} and \citet{Barreira2018response_approach}.

\dav{We quantify the severity of the impact with the ratio $\sigma_{\rm GS}/\sigma_{\rm G}$ between the marginalised or conditional uncertainties with and without SSC; this is found to vary significantly between different parameters and probes, and between the number and type of free nuisance parameters, in agreement with recent results \citep{upham2021, Barreira2018cosmic_shear, lacasa_20}}. \\
\dav{The conditional uncertainties show WL to be dramatically impacted by SSC, with all cosmological parameter uncertainties increasing by up to 210\% (for $\Omega_{{\rm m,0}}$) in the optimistic case. The GCph constraints are less sensitive to the addition of SSC, showing a smaller broadening of the uncertainties for all parameters, namely by up to one order of magnitude with respect to WL. The 3$\times$2pt case sits in between these two, while being in general closer to the GCph results because of its larger constraining power. \\
When considering marginalised constraints, the relative impact of SSC decreases significantly. In this case, the most impacted parameters for the single probes are mainly $\Omega_{{\rm m,0}}$, $w_0$, $w_a,$ and $\sigma_8$. Furthermore, the 3$\times$2pt becomes by far the most impacted probe, precisely because of its power in breaking parameter degeneracies. Indeed, in the reference case and the optimistic scenario, $({\rm FoM}_{w0wa}, {\rm FoM}_{\Omega_{\rm m, 0}S_8})$ decrease by (58\%, 41\%), hinting at the necessity to include SSC in the upcoming \Euclid analysis.} 

These results are the consequence of a complicated interplay between three factors. First, SSC originates from the uncertainty in the determination of the background mean density when measuring it over a finite region. This prevents determination of the overall amplitude of the matter power spectrum, which increases the uncertainty on those parameters that concur in setting its amplitude, mainly $\Omega_{{\rm m,0}}$ and $\sigma_8$. Second, the elements of the SSC matrix depend on the amplitude of the response functions. Third, the impact depends on how large a contribution the signal receives from the low-$z$ region, where the effective volume probed is smaller, making the variance of the background modes larger. The last two factors are both more severe for WL than for GCph, causing the former probe to be more affected than the latter. 

Finally, the deviation of a given element of the GS FM from the Gaussian one also depends on its correlations: in other words, the degradation of the constraints on a given parameter can be large if this is strongly correlated with a parameter severely degraded by SSC. Quantifying the impact of SSC on a single parameter is therefore quite hard in general, and must be investigated on a case-by-case basis taking care of the details of the probe and the way it depends on the parameter of interest.

Nuisance parameters to be marginalised over act as a sort of additional contribution to the covariance. As such, the importance of both the Gaussian and SSC contribution to the overall effective covariance becomes less important when the number of nuisance parameters increases. In order to consider cases that most closely mimic future \Euclid data, we opened up the parameter space by adding $\Omega_{{\rm DE,0}}$ \dav{(i.e. removing the flatness prior).} We find that, as long as the additional parameters have a scale-independent degeneracy with the most impacted ones, the relative impact of SSC decreases. We stress, however, that this reduction in the SSC impact has undesired consequences; the marginalised uncertainties on the parameters are definitely worsened, but the degradation is roughly the same whether the SSC is included or not, hence making the ratio $\sigma_{\rm GS}/\sigma_{\rm G}$ closer to unity for all parameters and probes. This result can be taken as a warning against investing too much effort in refining the estimate of the computationally expensive SSC when no approximations are made. For a \Euclid-like survey, the main concern would indeed be the number of nuisance parameters, which makes the impact of the SSC itself  less relevant.

We furthermore note that, in light of the recent theoretical developments presented in \citet{Lacasa2022}, it appears feasible to include the effect of SSC in the form of nuisance parameters, which would be the value of the density background $\delta_\mathrm{b}$ in each redshift bin. This approach is interesting as it would reduce the complexity of the data covariance matrix and would allow a simpler interpretation of the effect of SSC and how it is correlated to the other cosmological and nuisance parameters. 

Variations in the $z$ binning strategy have contrasting effects: a larger number of bins means a larger number of nuisance parameters (either galaxy bias or multiplicative shear bias for each bin), which leads to a loss of constraining power. Moreover, the larger the number of bins, the larger the Gaussian contribution to the covariance, making the shot and shape noise dominate over SSC for diagonal elements. On the downside, a larger number of bins leads to larger data vectors, thus adding information that can partially compensate for the increase in the covariance. The contrasting effects at play conspire in such a way that the degradation of the FoM due to SSC is found to be approximately independent of the number of redshift bins (cfr. Fig.~\ref{fig: ratioeped}). 

An interesting development in this sense is to leverage the SSC dependence on the low-$z$ contribution to investigate whether or not its impact could be mitigated by the use of the BNT transform, which transforms redshift bins in such a way as to increase the separation between the WL kernels. This will be investigated in a forthcoming work. 

An alternative strategy is to increase the constraining power by adding information through informative priors, hence recovering the FoM when SSC is incorrectly neglected. We investigate this possibility by quantifying the requirements on the prior information needed to recover the Gaussian FoM. Our results show that the main role is played here by the priors on galaxy bias parameters, while the FoM recovery is less sensitive to the prior on the multiplicative shear bias. However, the galaxy bias must be known to subpercent level in order to recover \dav{$\sim 70\%$} of the Gaussian FoM. Investigating whether or not this is possible is outside the scope of this paper. We nevertheless note that such remarkable prior information is the same as stating we are able to model the evolution of the bias with redshift. This is actually quite difficult based on the current knowledge of galaxy formation processes. Alternatively, one could look for an empirical fitting formula as a compromise between the need for strong priors on bias and the number of nuisance parameters.

\davtwo{One part of the covariance modelling not investigated in this work is the geometry of the survey footprint. While it is true that, for the large sky coverage considered, the full-sky approximation for SSC has been shown to suffice \citep{beauchamps2021}, a more realistic treatment accounting for the survey geometry should be considered for the Gaussian term. Still, we expect the main conclusions of this study to hold, as the mode coupling caused by the convolution with the survey mask, generating off-diagonal elements also in the Gaussian covariance, will mainly affect large scales where SSC is subdominant. It is however important to note that small holes in the survey mask (e.g. due to the presence of bright stars) generate mode coupling also on small scales, where we have seen the SSC impact to be most prominent. This will be investigated in future works.}

Although some more work is needed to improve the robustness of our results, for example by comparing the different approximations presented in the literature, we can conclude that the effect of including the SSC term in the total covariance matrix of \Euclid photometric observables is definitely non-negligible, especially for WL and 3$\times$2pt. However, the degradation of the constraints on cosmological parameters depends on the particular probe and the number and kind of parameters to constrain. The FoM is nevertheless reduced by \dav{}{$52\%$ ($45\%$)} for the 3$\times$2pt probe in the pessimistic (optimistic) scenario where all cosmological (including $\Omega_{\rm DE,0}$) and nuisance (multiplicative shear bias) parameters are left free to vary. Maximising the power of the actual \Euclid photometric data by taking into account the presence of SSC is a daunting task, which we will report on in a forthcoming publication.\\

%%%%%%%%%%%%%%%%%%%%%%%%%%%%%%%%%%%%%%%%%%%%%%%%%%%%%%%%%%%%%%%%%%%%%%%

\begin{acknowledgements}
The computational part of the work has been performed using the \texttt{Python} programming language, interfaced with scientific packages like \texttt{astropy} 
\citep{astropy2013, astropy2018} for cosmological calculations, \texttt{Numba} \citep{numba} for code speedup, \texttt{NumPy} \citep{harris2020array} for matrix manipulation, \texttt{SciPy} \citep{2020SciPy-NMeth} for numerical integration and \texttt{Matplotlib} \citep{Hunter:2007} for data visualisation. 
The authors thank the anonymous referees for their helpful comments that improved the quality of the manuscript. DS would like to thank Raphael Kou for the fruitful discussion on the SSC impact on GCph.  SGB was supported by CNES, focused on \Euclid mission. The project leading to this publication has received funding from Excellence Initiative of Aix-Marseille University -A*MIDEX, a French "Investissements d'Avenir" programme (AMX-19-IET-008 -IPhU). SC acknowledges support from the `Departments of Excellence 2018-2022' Grant (L.\ 232/2016) awarded by the Italian Ministry of University and Research (\textsc{mur}). IT acknowledges funding from the European Research Council (ERC) under the European Union's Horizon 2020 research and innovation programme (Grant agreement No.\ 863929; project title ``Testing the law of gravity with novel large-scale structure observables'' and acknowledges support from the Spanish Ministry of Science, Innovation and Universities through grant ESP2017-89838, and the H2020 programme of the European Commission through grant 776247. 
\AckEC
\end{acknowledgements}

%We thank XXX for providing YYY, as well as an anonymous referee for constructive comments. 

%%%%%%%%%%%%%%%%%%%%%%%%%%%%%%%%%%%%%%%%%%%%%%%%%%%%%%%%%%%%%%%%%%%%%%%

\bibliographystyle{aa} % shortens the references list!
\bibliography{main} % Entries are in the &quot;refs.bib&quot; file</code></pre>

%%%%%%%%%%%%%%%%%%%%%%%%%%%%%%%%%%%%%%%%%%%%%%%%%%%%%%%%%%%%%%%%%%%%%%%

\begin{appendix}

\section{Details of the code validation} \label{sec: validation_appendix}

In the following, we provide an overview of the steps undertaken to compare and validate the codes used in this work, and some of the lessons learnt in the process.

\begin{figure}%[h]
\includegraphics[width=\hsize]{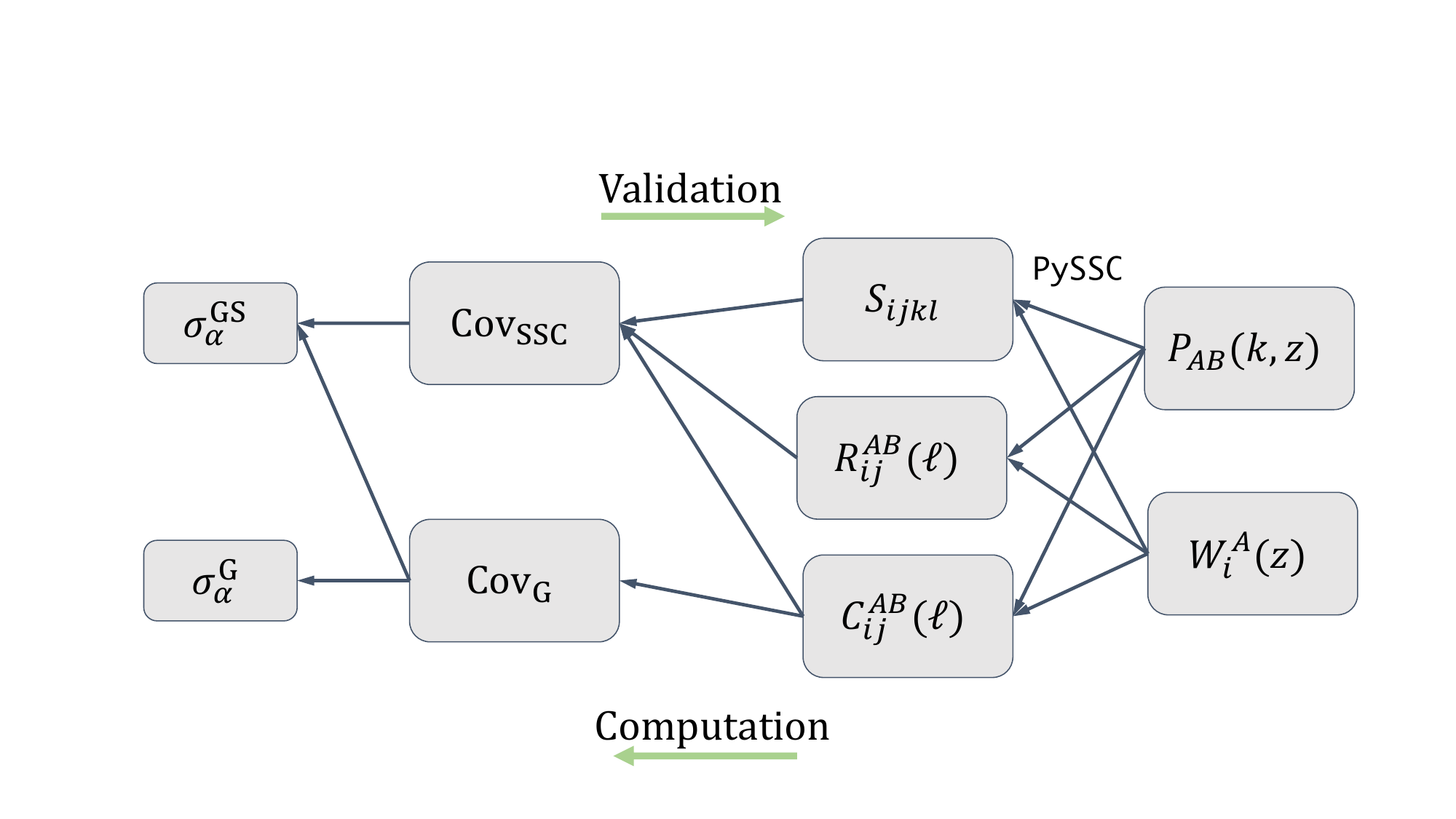}
\caption{Some of the most important elements examined in the comparison. The arrows show the ordering followed to produce the parameters constraints, which is opposite to the one followed to validate the code. The derivatives of the PS with respect to the cosmological parameters, entering the final step of the computation, are not shown.}
\label{fig: procedure}
\end{figure}

In order to compute and validate the results we adopt the scheme sketched in Fig.~\ref{fig: procedure}, which highlights the dependency of each main element of the forecast computation on the others. In particular, we have that:
\begin{enumerate}
    \item The $1\sigma$ constraints are obtained from the FM through Eq.~\eqref{eq: cramer_rao}, and the FM is built in turn from the (inverse) covariance matrix and the derivatives of the angular PS $C_{ij}^{AB}(\ell)$ as indicated in Eq.~\eqref{eq: fishmat}.
    \item The Gaussian covariance depends on the $C_{ij}^{AB}(\ell)$ through Eq.~\eqref{eq: covgauss} (and the noise PS, Eq.~\ref{eq: noiseps}). The SSC also depends on the $C_{ij}^{AB}(\ell)$, with the added contribution of the $R_{ij}^{AB}(\ell)$ terms and the output of the \texttt{PySSC} module, the $S_{ijkl}$ matrix -- following Eq.~\eqref{eq: covssc}. 
    \item The $C_{ij}^{AB}(\ell)$ are constructed by convolving the (non-linear) matter PS with the lensing and galaxy weight functions, as in Eq.~\eqref{eq: CijSSC}. The $S_{ijkl}$ matrix also depends on the weight functions (see Eq.~\ref{eq: sijkl}), which are in fact the main external input needed by \texttt{PySSC}, and on the \textit{linear} matter PS through the $ \sigma^{2}(z_{1}, z_{2})$ term (Eq.~\ref{eq: sigma}). It is to be noted, however, that \texttt{PySSC} computes this PS internally, needing only the specification of a dictionary of cosmological parameters with which to call the Boltzmann solver \texttt{CLASS} through the  \texttt{Python} wrapper \texttt{classy}. This means that we also have to make sure that the fiducial value of the parameters used to compute the PS of Eq.~\eqref{eq: CijSSC} are the same ones passed to \texttt{PySSC} (this time to compute the linear PS), in order to work with the same cosmology.
\end{enumerate}
While to compute the constraints we follow the scheme from right to left, starting from the basic ingredients to arrive at the final result, the general trend of the validation is the opposite: we begin by comparing the final results, then work our way back whenever we find disagreement.

We then start the comparison from the $\sigma_\alpha$. If a discrepancy larger than $10\%$ is found, we check the quantities they depend on, which in this case are the covariance matrices (see Eq.~\ref{eq: fishmat}). If these agree, we check the codes directly. If these disagree, we iterate the process by checking the subsequent element in the scheme (in this case the $S_{ijkl}$ matrix and the $C_{ij}^{AB}(\ell)$), until agreement is found. Essentially, this means that the disagreement in the outputs of the codes at each step can either come from the inputs, or from the codes themselves. Once the cause of the discrepancy is found and fixed, the computation is repeated and the process can start again.

The pipelines under comparison are both written in the $\texttt{Python}$ language. One of them requires as external inputs the weight functions, the angular PS $C^{AB}_{ij}(\ell)$ and their derivatives with respect to the cosmological parameters; whilst the other produces these through the use of $\texttt{CosmoSIS}$\footnote{\texttt{\url{https://bitbucket.org/joezuntz/cosmosis/wiki/Home}}} \citep{Zuntz_2015}, and hence needs no external inputs but the vectors of fiducial cosmological and nuisance parameters.
For the reader wishing to repeat the validation, we list below some of the lessons learnt in the code comparison process.
\begin{itemize}
    \item \texttt{PySSC} needs as input the WL and GCph kernels of Eqs.~\eqref{eq: wildef} and \eqref{eq: wigdef}, as well as their argument, the redshift values. The code then uses this redshift array to perform the necessary integrals in $\diff V$ through Simpson's rule. The user is responsible for sampling the kernels on a sufficiently fine $z$ grid [$O(10^4)$ values have been found to be sufficient in the present case] to make sure these integrals are performed accurately.
    \item The latest version of \texttt{PySSC} accepts a \texttt{convention} parameter. This specifies whether the kernels are in the form used in \citetalias{Lacasa_2019} (\texttt{convention = 0}) or the one prescribed in \citetalias{ISTF2020} (\texttt{convention = 1}). The two differ by a $1/r^2(z)$ factor, as shown in Eq.~\eqref{eq: wfmatch}. Passing the kernels in the \citetalias{ISTF2020} form without changing the parameter's value from \texttt{0} -- the default -- to \texttt{1} will obviously yield incorrect results. 
    \item The ordering of the $S_{ijkl}$ matrix's elements depends on the ordering chosen when passing the input kernels to \texttt{PySSC}  -- whether $W_i^{\rm L}(z)$ first and $W_i^{\rm G}(z)$ second or vice versa. This must be kept in mind when implementing Eq.~\eqref{eq: covssc}.
    \item The GCph constraints can show a discrepancy greater the 10\% for the dark energy equation of state parameters $w_0$ and $w_a$ even when the corresponding covariance is found to be in good agreement. This discrepancy is due to GCph being less numerically stable because of the lower constraining power compared to the other probes, and because the bias model considered has a strong degeneracy with $\sigma_8$, making the numerical derivatives unstable \citep[see e.g.][]{Casas2023}. Since this is a known issue, not coming from the SSC computation, and the covariance matrices and angular PS show good agreement,
    % XXX and the main aim of this work is to validate the computation of the Super Sample \textit{Covariance} (albeit through its impact on the parameter constraints)
    we choose to overcome the problem by using, for GCph, one code to compute both sets of parameter constraints (that is, we run one FM evaluation code with as input the covariance matrices from both groups).
\end{itemize}
\section{High order bias from halo model} \label{sec: halomodel}

As described in Sect.~\ref{sec: higher_order_bias}, the higher-order bias $b_{(2)}(z)$ has been estimated using the halo model. In the following, we provide further details on the input quantities, and how we set the relevant parameters. 

A key role is played by the halo mass function $\Phi_{\rm MF}(M, z)$, which we model as 
\begin{equation}
\Phi_{\rm MF}(M, z) = \frac{\bar\rho_{\rm m}}{M} f(\nu) \frac{\diff\ln{\sigma^{-1}}}{\diff M},
\label{eq: hmft10}
\end{equation}
with $M$ the halo mass, ${\bar\rho}_{\rm m}$ the mean matter density, $\nu= \delta_{\rm c}/\sigma(M, z)$, $\delta_{\rm c} = 1.686$ the critical overdensity for collapse, and $\sigma(M, z)$ the variance of linear perturbation smoothed with a top-hat filter of radius $R = \left[3M/(4 \pi \bar\rho_{\rm m})\right]^{1/3}$. We follow \citet{Tinker2010}, setting 
\begin{equation}
f(\nu) = {\cal N}_{\rm MF} \left [1 + \left ( \beta_{\rm MF} \nu \right )^{\, -2 \phi_{\rm MF}} \right ] \nu^{\, 2 \eta_{\rm MF}} \exp{\left ( - \gamma_{\rm MF} \nu^{\, 2}/2 \right )},
\label{eq: fnumf}
\end{equation}
where ${\cal N}_{\rm MF}$ is a normalisation constant, and the halo mass function fitting parameters $\beta_{\rm MF}, \eta_{\rm MF}, \gamma_{\rm MF}$ and $\phi_{\rm MF}$ -- not to be confused with $\Phi_{\rm MF}(M, z)$ -- scale with redshift as illustrated in Eqs.~(9--13) of the above-mentioned paper.

The other quantity needed is the average number of galaxies hosted by a halo of mass $M$ at redshift $z$. This is given by 
\begin{equation}
\langle N|M \rangle(M) = 
N_{\rm cen}(M) \left[1 + N_{\rm sat}(M)\right] \; ,
\label{eq: nmhod}
\end{equation}
where $N_{\rm cen}(M, z)$ and $N_{\rm sat}(M, z)$ account for the contributions of central and satellite galaxies, respectively. We model these terms as in \citet{White2011}
\begin{equation}
N_{\rm cen}(M) = \frac{1}{2} \left \{ 1 +  {\rm erfc}{\left [\frac{\ln{\big(M/M_{\rm cut}\big)}}{\sqrt{2} \sigma_c} \right ]} \right \} \; ,
\label{eq: ncen}
\end{equation}
\begin{equation}
N_{\rm sat}(M) = \left \{
\begin{array}{ll}
\displaystyle{0} & \displaystyle{M < \kappa_s M_{\rm cut}} \\
 & \\
\displaystyle{\left ( \frac{M - \kappa_s M_{\rm cut}}{M_1} \right)^{\alpha_s}} & \displaystyle{M \ge \kappa_s M_{\rm cut}}, \\
 \end{array}
\right . 
\label{eq: nsat}
\end{equation}
with fiducial parameter values
\begin{align}
\{\logten{(M_{\rm cut}/M_\odot)}, &\logten{(M_1/M_\odot)}, \sigma_c, \kappa_s, \alpha_s \} \nonumber \\
& = {13.04, 14.05, 0.94, 0.93, 0.97} \; ,
\end{align}
$M_\odot$ being the mass of the Sun. These values give the best fit to the clustering of massive galaxies at $z \sim 0.5$ as measured from the first semester of BOSS data. It is, however, expected that they are redshift-dependent although the precise scaling with $z$ also depends on the galaxy population used as a tracer. We therefore adjust them so that the predicted galaxy bias matches, at each given redshift, our measured values from the Flagship simulation. Since, for each $z$, we have a single observable quantity, we cannot fit all parameters. On the contrary, we fix all of them but $M_{\rm cut}$ to their fiducial values and use Eq.~\eqref{eq: bicalc} to compute the bias as a function of $M_{\rm cut}$. We then solve with respect to $M_{\rm cut}$ repeating this procedure for each redshift bin. We then linearly interpolate these values to get $M_{\rm cut}$ as a function of $z$, and use it to compute $b_{(2)}(z)$. Although quite crude, we have verified that changing the HOD parameter to be adjusted (e.g. using $\sigma_c$ or $M_1$) has a negligible impact on the predicted $R^{\rm gm}(\ell)$ and $R^{\rm gg}(\ell)$.

\section{Multipole binning}\label{sec: appendix_binning}
We bin the $\ell$ space  according to the following procedure: the $\ell_k$ values, where $k = 1, ..., {\cal N}_\ell$, are the centres of ${\cal N}_\ell + 1$ logarithmically equispaced values, $\lambda_{k}$, which act as the edges of the ${\cal N}_\ell$ bins:
\begin{equation}
\ell_k = {\rm dex}\brackets{
    \paren{\lambda_{k}^{-} + \lambda_{k}^{+}}/2
    } \; ,
\label{eq: ellkxc}
\end{equation}
with ${\rm dex}(x) = 10^x$, $\paren{\lambda_{k}^{-}, \lambda_{k}^{+} } = \paren{\lambda_{k}, \lambda_{k + 1}}$, and
\begin{equation}
\lambda_k = \lambda_{\rm min}^{\rm XC} + (k - 1) (\lambda_{\rm max}^{\rm XC} - \lambda_{\rm min}^{\rm XC})/{\cal N}_{\ell} \; ,
\label{eq: lambdakxc}
\end{equation}
being
\begin{equation}
\left\{\lambda_{\rm min}^{\rm XC}, \lambda_{\rm max}^{\rm XC}\right\} = \left\{
\logten{(\ell_{\rm min}^{\rm XC})}, \, \logten{(\ell_{\rm max}^{\rm XC})}\right\} \; .
\label{eq: rangexc}
\end{equation}
In order to compute the Gaussian covariance, we also need the width of the bin, which will simply be
\begin{equation}
\Delta \ell_k = {\rm dex}(\lambda_{k + 1}) - {\rm dex}(\lambda_k) \; ,
\label{eq: deltaell} 
\end{equation}
so that $\Delta \ell_k$ is not the same for all bins, since the bins are logarithmically -- and not linearly -- equispaced. 

\end{appendix}

\end{document}